\documentclass[12pt,letterpaper,usenames,dvipsnames]{article}
\usepackage{Auxiliary/jheppub}

\usepackage{amsmath,amssymb,amsfonts,amsthm,amscd,setspace}
\allowdisplaybreaks[1]
\usepackage{bm} 
\usepackage{bbm} 
\usepackage{mathbbol} 
\usepackage{mathrsfs} 

\usepackage{graphicx,tikz}
\usepackage[framemethod=TikZ]{mdframed} 
\usepackage{tikz-cd} 
\usetikzlibrary{arrows,decorations,shapes.arrows,decorations.markings}
\tikzset{>=triangle 90}
\tikzstyle{g}=[draw,circle,green!50!black,fill=green!50!black,scale=.6]
\tikzstyle{R}=[draw,circle,fill=red,scale=.6]

\usepackage{multicol}
\usepackage[export]{adjustbox}
\usepackage{rotating} 
\usepackage{colortbl}
\usepackage{tcolorbox,array,hhline,arydshln,multirow,floatrow,subfig}

\usepackage{enumitem}

\usepackage{xcolor}

\def\blue#1{{\color{blue}{#1}}}

\usepackage{comment}

\def\nn{\nonumber}
\def\ph{\phantom}
\def\bM{\begin{matrix}}
\def\eM{\end{matrix}}
\newcommand{\bpm}{\begin{pmatrix}}
\newcommand{\epm}{\end{pmatrix}}
\newcommand{\bsm}{\begin{smallmatrix}}
\newcommand{\esm}{\end{smallmatrix}}
\newcommand{\bspm}{\left(\begin{smallmatrix}}
\newcommand{\espm}{\end{smallmatrix}\right)}
\def\bar{\overline}
\def\til{\widetilde}

\def\^{\wedge}
\def\del{{\partial}}

\def\vev#1{{\langle{#1}\rangle}}

\def\Im{{\rm Im}}

\def\im{{\rm im}}

\def\aut{{\rm aut}}

\def\CB{{\rm CB}}
\def\GL{{\rm GL}}

\def\SL{{\rm SL}}
\def\PSL{{\rm PSL}}

\def\SU{{\rm SU}}
\def\su{\mathfrak{su}}
\def\Sp{{\rm Sp}}

\def\U{{\rm U}}
\def\u{\mathfrak{u}}
\def\cA{{\mathcal A}}
\def\cB{{\mathcal B}}
\def\C{\mathbb{C}}

\def\cF{{\mathcal F}}
\def\fg{{\mathfrak g}} 
\def\cH{{\mathcal H}}
\def\cM{{\mathcal M}}
 
\def\cN{{\mathcal N}}

\def\P{\mathbb{P}} 
\def\Q{\mathbb{Q}} 
\def\R{\mathbb{R}}

\def\cV{{\mathcal V}}

\def\Z{\mathbb{Z}}
\def\a{{\alpha}}

\def\b{{\beta}}
\def\g{{\gamma}}

\def\d{{\delta}}
\def\D{{\Delta}}

\def\z{{\zeta}}

\def\l{{\lambda}}

\def\r{{\rho}}
\def\s{{\sigma}}

\def\S{{\Sigma}}
\def\t{{\tau}}

\def\w{{\omega}}
\def\Om{{\Omega}}

\def\suf{\mathfrak{su}}

\def\cA{{\mathcal A}}
\def\cB{{\mathcal B}}

\def\cF{{\mathcal F}}

\def\cH{{\mathcal H}}

\def\cM{{\mathcal M}}

\def\cN{{\mathcal N}}

\def\cV{{\mathcal V}}

\def\P{\mathbb{P}}
\def\Q{\mathbb{Q}}  
\def\R{\mathbb{R}} 
 
\def\Z{\mathbb{Z}} 

\def\beq{\begin{equation}}
\def\eeq{\end{equation}}
\def\nn{\nonumber}
\newcommand{\bpmat}{\begin{pmatrix}}
\newcommand{\epmat}{\end{pmatrix}}
\newcommand{\bsmat}{\begin{smallmatrix}}
\newcommand{\esmat}{\end{smallmatrix}}
\newfloatcommand{capbtabbox}{table}[][\FBwidth]

\title{The rank-2 classification problem III:\\ 
curves with additional automorphisms}
\author[1]{Philip C. Argyres}
\author[2]{Mario Martone}
\affiliation[1]{University of Cincinnati, Physics Department, PO Box 210011, Cincinnati OH 45221}
\affiliation[b]{Dept.\ of Mathematics, King’s College London, The Strand, London WC2R 2LS, UK}
\emailAdd{philip.argyres@gmail.com}
\emailAdd{mario.martone@kcl.ac.uk}

\abstract{This is the third in a series of papers which outlines an approach to the classification of $\cN{=}2$ superconformal field theories at rank 2 via the study of their Coulomb branch geometries. 
Here we use the fact that the encoding of a Coulomb branch geometry as a Seiberg-Witten curve and 1-form enjoys a large reparametrisation invariance.
While there is always a unique way to fix this invariance such that the curve and 1-form are single-valued over the Coulomb branch --- the ``canonical frame" of the curve used in the first two papers in this series --- there are other useful frames in which the curve is single-valued but the 1-form is allowed to be multi-valued.
In these frames, which we call ``automorphism frames", the 1-form is periodic up to an automorphism twist.
We argue that the multi-valuedness of the automorphism frame can  simplify the computational complexity of finding new consistent scale invariant solutions. 
We demonstrate this in an example by using the automorphism frame to construct for the first time a genus 2 Seiberg-Witten curve for the $\cN{=}4$ $\su(3)$ superYang-Mills theory, a solution that is hard to find by other approaches.}
  
\begin{document}
\maketitle 

\section{Introduction}

The aim of classifying all possible rank-2 scale-invariant special K\"ahler (SK) Coulomb branch (CB) geometries is motivated by the constraints such a classification would put on the set of possible 4d $\cN{=}2$ superconformal field theories (SCFTs).
Also, new examples of such geometries would be indications of previously unknown SCFTs. We dedicate a short series of three papers, comprising this one and \cite{Argyres:2022lah,AM22II}, to setting up a viable way of achieving this goal.

Throughout the series, we simplify our task by making the assumptions that the Dirac pairing is principal and that the CB $\simeq \C^2$ as a complex space.
These are not physically benign assumptions since counter-examples are known \cite{Bourget:2018ond, Argyres:2018wxu, Caorsi:2018zsq, Argyres:2022kon}.
Even with these assumptions, the CB geometry is non-trivial: there will be a complex codimension 1 subvariety, which we call the \emph{singular locus} $\cV\subset\C^2$, where the CB metric is non-analytic.
The non-singular open submanifold of the CB is $\CB^* \doteq \CB \setminus \cV$.

With these assumptions, a general rank $2$ CB geometry can be written in terms of complex geometry as a family of genus $2$ Riemann surfaces (RSs) varying holomorphically over CB$^*$ together with a choice of basis of holomorphic 1-forms on the RSs which vary locally holomorphically on the CB. 
The family of RSs and 1-form bases must satisfy certain conditions imposed by unitarity and superconformal invariance in order to describe a SK geometry and we refer to this data as the Seiberg-Witten (SW) data of the CB.

Furthermore, all non-singular genus 2 RSs can be written as a suitable projectivization%
\footnote{The suitable projectivization is $y^2 = c(x,w)$ where $c$ is homogeneous of degree 6 in $x$ and $w$ and $[w:x:y] \in \P^2_{(1,1,3)}$ are homogeneous coordinates of a weighted projective space.}
of binary sextic plane curves,
\begin{align}\label{SWcurve}
    y^2 &= c(x;u,v), 
\end{align}
where $c$ is a degree-6 polynomial in $x$ and $(u,v) \in \C^2$ are complex coordinates on the CB.
The SW curve is then a family of such genus 2 curves together with a basis of 1-forms varying locally holomorphically over $\CB^*$.
So, at least with the above assumptions, there is a uniform and relatively simple algebraic setting in which rank 2 CB geometries can be explored.

This setting was used in a previous attempt \cite{Argyres:2005pp, Argyres:2005wx} at systematically constructing scale-invariant rank 2 CB geometries by finding all $c(x,u,v)$ in \eqref{SWcurve} meromorphic in $(u,v)$ satisfying the condition following from unitarity and scale invariance --- \emph{i.e.} the SK integrability condition reviewed in appendix \ref{appA}. 
In the previous two papers in this series \cite{Argyres:2022lah, AM22II} we improved the analysis in \cite{Argyres:2005pp, Argyres:2005wx} by lifting a physically unmotivated assumption and by developing techniques to analyse mass deformations of scale invariant geometries.
This attempt was also incomplete, for though it extended the previous list of curves, it still did not find curves corresponding to the known rank-2 $\cN=3$ and 4 SCFTs. 
This is because in order to define a problem which we could solve completely, we made another simplifying, yet physically unmotivated, assumption: we restricted to polynomial coefficients in the curve \eqref{SWcurve} \cite{Argyres:2022lah}. 
Extending the work of \cite{Argyres:2022lah, AM22II} beyond polynomial coefficients seems currently computationally unfeasible.

A way to proceed, which will be the focus of this paper, is to follow a new approach to constructing CB geometries which can be computationally advantageous in the more general case of non-polynomial coefficients. 
This approach is based on the observation that the aforementioned integrability conditions only constrain the SW data to be \emph{locally} holomorphic on CB$^*$. In particular, curve and 1-form coefficients may be branched over $\cV \subset \CB$, and so be multi-valued on $\CB^*$ and this can be used to simplify the search for non-polynomial solutions.
By contrast, the analysis in \cite{Argyres:2005pp, Argyres:2005wx, Argyres:2022lah, AM22II} only considered the case of SW data with trivial global properties: the coefficients of $c(x,u,v)$ were meromorphic in $(u,v)$ and the choice of the 1-form bases constant over the CB. 
Since the algebraic description of the curve and 1-form basis enjoys a continuous reparameterization freedom, the global behaviour of the SW data is not an invariant and the same CB can be described by many different sets of SW data. 
We will show that this freedom can always be fixed in such a way that the 1-form basis takes a canonical form where the curve is meromorphic (\emph{i.e.}, single-valued) over $\CB^*$. 
We call this way of fixing the reparameterization invariance --- the framework used in nearly all previous works on the subject --- the \emph{canonical frame} for the SW data.  
In this sense the approach in \cite{Argyres:2022lah, AM22II}, once the polynomial ansatz is lifted, is general.

Here we propose fixing the reparameterization invariance in a different manner, which we call the \emph{automorphism frame}, where the coefficients of the 1-form basis are not single valued.
We will show that such frames often lead to algebraically simple solutions of the SK integrability conditions which are instead complicated once expressed in the canonical frame. 
The reason for the name ``automorphism frame" is due to the fact, which will be shown below, that the physically allowed multi-valuedness is closely tied to a representation of the fundamental group of $\CB^*$ into the automorphism group of the family of RSs. Since $\pi_1(\CB^*)$ is a finitely generated discrete group and the automorphism groups of genus-2 RSs are finite groups, this representation is some discrete topological part of the SW data which we call the \emph{automorphism twist} of the SW data. 
We will see that a choice of automorphism frame leaves unfixed a finite subgroup of reparametrizations, namely the automorphism group of the RS, where the unfixed automorphism group enters as periodicity conditions on the 1-form basis coefficient matrix, reflecting the automorphism twist of the data.

Finally, we demonstrate that, at least in a large class of CB geometries, the SK integrability conditions can be solved in terms of a single polynomial equation.
We show in the example of the $\cN{=}4$ $\su(3)$ sYM theory how the polynomial in question is determined by the automorphism twist data.
This allows us to derive, for the first time, a genus 2 SW curve for the $\cN{=}4$ $\su(3)$ sYM theory; this curve was previously announced in \cite{Argyres:2022kon}.%
\footnote{A genus 3 curve for the $\cN{=}4$ $\u(3)$ sYM theory has been known for a long time \cite{Donagi:1995cf}, which must be supplemented by an additional condition to describe the $\su(3)$ theory.  It turns out that that description --- genus 3 curve plus decoupling condition --- describes the theory with a non-principal Dirac pairing.  See \cite{Argyres:2022kon} for a discussion.}


We introduce the automorphism twist of the SW data in section \ref{sec3}, after briefly reviewing the conditions coming from $\cN{=}2$ supersymmetry, scale-invariance, and unitarity that genus 2 SW curves must satisfy.
Section \ref{secBS} describes genus 2 RSs with extra automorphisms, introduces the canonical and automorphism frames, and discusses how this can be used to construct SW curves for the special class of ``isotrivial" CB geometries.
Section \ref{sec4} applies this setup to a concrete example and uses the automorphism frame to construct for the first time the SW data for the $\su(3)$ $\cN{=}4$ theory.
Section \ref{sec5} proposes a classification program for rank 2 CB geometries based on their curve automorphisms, and discusses some of its main challenges.
In appendix \ref{appA} we review the integrability conditions while the remaining three appendices collect and derive relevant facts about the automorphisms of genus 1 and genus 2 RSs.

\section{Global properties of genus 2 SW curves --- the automorphism twist}
\label{sec3}

This section discusses the global conditions allowed on SW data, specifically the allowed monodromies for the 1-form basis and the basis of 1-cycles on the RS fibered on a generic point of the CB. Before doing that, let's briefly review the basics of rank 2 CB geometry.

We assume that, as a complex space, the CB $\simeq_\C \C^2$; in particular, it has no complex singularities.  
Choose complex coordinates, $(u_1,u_2) \doteq (u,v) \in \C^2$, unique up to normalization, which diagonalize the holomorphic $\C^*$ symmetry action enjoyed by scale-invariant CB geometries,
\begin{align}\label{Cstar}
\C^* : (u,v) &\mapsto (t^{\D_u} u, t^{\D_v} v) , &
t &\in\C^*, & \D_u \le \D_v .
\end{align}
$\D_j$ are the CB scaling dimensions which are positive rational numbers \cite{Argyres:2018zay, Caorsi:2018zsq, Argyres:2018urp}, and which we write as
\begin{align}\label{CBdims}
(\D_u,\D_v) &= (p,q) R & 
&\text{for} &
0&<R\in\Q, &
&\text{and}&
p,q &\in \Z, &
\gcd(p,q)&=1.
\end{align}
The finite set of allowed pairs $(\D_u,\D_v)$ is given by a refinement of an argument in \cite{Caorsi:2018zsq, Cecotti22}.

An SK structure with principal Dirac pairing on a scale-invariant CB can be encoded in a family of genus 2 RSs, $\S$, together with a basis of holomorphic 1-forms, $\Om_i$, varying locally holomorphically over the CB.%
\footnote{As explained in the introduction of \cite{Argyres:2022lah}, there is a special class of ``split" SK structures for which this is not true.  In that case the family of genus 2 RSs should be replaced by a family of pairs of elliptic curves with their marked points identified.  We will not discuss this family here.\label{split ftnt}}
A general genus 2 RS can be written as a binary sextic plane curve,
\begin{align}\label{SWcrv}
\S: \qquad
y^2 &= c(x,u,v) \doteq \sum_{n=0}^6 c_n(u,v) x^n ,
\end{align}
so the $c_n$ coefficients vary locally holomorphically over the CB.  
We define a \emph{canonical basis} of its holomorphic 1-forms to be
\begin{align}\label{1fbasis1}
\w_u &:= x \, \frac{dx}{y} ,&  \w_v &:= \frac{dx}{y} .
\end{align}
A general basis of holomorphic 1-forms is then
\begin{align}\label{1fbasis2}
\Om_i := \w_j \, a_i^j ,
\end{align}
where $a_i^j$ vary locally holomorphically over the CB.
The curve and basis of 1-forms determine an SK structure if they satisfy the \emph{SK integrability condition}
\begin{align}\label{SWint}
\del_v\Om_u - \del_u \Om_v = d g,
\end{align}
where $\del_i := \del/\del u_i$, $g$ is a meromorphic function on the RS, and $d$ is the exterior derivative on the RS. \eqref{SWint} gives rise to eight coupled ordinary differential equations, reviewed in detail in appendix \ref{appA}, which will be solved in section \ref{sec4} to construct for the first time the SW curve of the $\suf(3)$ $\cN=4$ theory.
For a scale-invariant CB, the central charge, $Z$, of the low energy supersymmetry algebra on the CB is given by the periods
\begin{align}\label{Zd}
    Z(\d) &= \oint_\d \Lambda, &
    \Lambda &\doteq (1+\D_x)^{-1}  (\D_u u\, \Om_u + \D_v v\, \Om_v), &
    \d &\in H_1(\S).
\end{align}
where $\D_x$ is the scaling dimension of $x$.
Here $\Lambda$ is the SW 1-form.
Special coordinates, metric, and matrix of electromagnetic couplings on the CB are then determined from the periods of the $\Om_i$ in the usual way, and will be reviewed where needed in the next subsection.

The metric non-analyticities on the CB form an algebraic subvariety, $\cV$, which we call the singularity locus of the CB.
By scale invariance, $\cV$ is a finite union of component sub-loci given by
\begin{align}\label{knots}
u &=0, & v &=0, & & \text{and/or} &
u^q &= t v^p, & t &\in \C^* . 
\end{align}
In case the basis of holomorphic 1-forms, $\Om_i$, is non-degenerate and non-singular everywhere on the CB, then the singular locus, $\cV$, is given by a subset of the set of zeros of the discriminant of $c(x)$ with respect to $x$,
\begin{align}\label{Ddef}
    \cV \subset \{ \text{Disc}_x(c)=0 \} \subset \text{CB},
\end{align}
because the zeros of the discriminant of the curve determine where cycles of the curve degenerate.  Only homologically non-trivial degenerating cycles give rise to non-analyticities in the CB SK metric.
But, as we will see below, the discriminant of the curve does not necessarily capture the singularity locus on the CB in the case where the basis $\Om_i$ of holomorphic 1-forms has poles or degenerations.

\subsection{Physically allowed multi-valuedness of the SW data}

The key to understanding the constraints on the possible multi-valuedness of the SW curve and 1-form basis on CB$^*$ is to identify the properties of the SW curve and 1-form basis which are physically observable, i.e., which enter into the description of the low energy physics on the CB. 

The value of the central charge \eqref{Zd} is a physical observable: its norm computes the mass of BPS particles, and its phase certain CP-violating phases in the presence of such particles.
The central charge is a function of the CB vacuum.  
Given a choice of scaling coordinates $(u,v)$ on the CB, which are generally uniquely determined up to an overall complex normalization,%
\footnote{In the case both $u$ and $v$ have the same scaling dimension, then they are unique only up to a $\GL(2,\C)$ linear reparameterization.}
then the CB derivatives of the central charge, 
\begin{align}\label{delZ}
    \del_i Z(\d) = \oint_\d \Om_i, 
\end{align}
is also an observable.
The central charge is also a function of the rank-4 lattice of $\U(1)^2$ EM charges on the CB.
This charge appears in \eqref{delZ} as a choice of homology 1-cycle $\d \in H_1(\S)$.
This charge lattice is endowed with a Dirac pairing, $J$, which we are assuming is principal, meaning there is a basis of homology 1-cycles $\{\a^i,\b_i\}$, $i=1,2$, such that $J(\a^i,\a^j)=J(\b_i,\b_j)=0$ and $J(\a^i,\b_j)=\d_j^i$.
We will call such a basis a \emph{canonical 1-cycle basis}, and refer to the $\a^i$ as ``electric" and the $\b_i$ as ``magnetic" charges.
The Dirac pairing corresponds to the intersection pairing on the homology 1-cycles.
Write the 4-vector of homology basis 1-cycles as
\begin{align}\label{1-cycle basis}
    \d_a \doteq \bpm \b_i \\ \a^i \epm .
\end{align}
(Note that we put the magnetic charges in the top block.)

\subsubsection{The 1-cycle basis monodromy}

Pick a canonical 1-cycle basis $\d_a$ at some base point $p\in \CB^*$. Then define the 4-vector \emph{special section}, $\s_a$, and the $4\times2$ \emph{period matrix} of the 1-form basis, $\Pi_{a\,j}$, by
\begin{align}\label{periodmat}
    \s_a &\doteq \oint_{\d_a} \Lambda ,&
    \Pi_{a\,j} &\doteq \oint_{\d_a} \Om_j .
\end{align}
These are functions both of the CB vacuum, $p\in\CB$, and the choice of 1-cycle basis, $\d_a$, at $p$.
They both depend locally holomorphically on $p\in\CB^*$.
Relative to a choice of canonical 1-cycle basis, decompose $\s$ and $\Pi$ into magnetic and electric blocks,
\begin{align}\label{specialcoords}
    \s_a &\doteq \bpm s^D_i \\ s^i \epm, &
    \Pi_{a\,j} &\doteq \bpm \cB_{ij} \\ \cA^i_j \epm
    = \bpm \del_j s^D_i \\ \del_j s^i \epm ,
\end{align}
where $s^i$ and $s_i^D$ are the vectors of \emph{special coordinates} and dual special coordinates with $i=1,2$.
Then the matrix of electromagnetic couplings on the CB is
\begin{align}\label{taudef}
    \t_{ik} &\doteq \cB_{ij} (\cA^{-1})^j_k.
\end{align}
It follows from unitarity of the effective action on the CB, and equally from the Riemann bilinear relations on the RS, that $\t_{ij}$ is a $2\times2$ symmetric complex matrix with positive-definite imaginary part.
These conditions define the rank 2 Seigel half-space, $\cH_2$; it is a (non-compact) 3-complex-dimensional manifold.%
\footnote{``Split" degenerate genus 2 RSs, mentioned in footnote \ref{split ftnt}, are included as a 2-complex-dimensional subvariety of $\cH_2$.}
Note that, by \eqref{taudef}, the period matrix can be written in terms of $2\times 2$ blocks as
\begin{align}\label{Pitotau}
    \Pi = \bpm \t \\ 1 \epm \cA .
\end{align}
$\t$ is independent of the choice of 1-form basis $\Om_j$, and encodes the complex structure modulus of the RS.

The basis of 1-cycles $\d_a$ can be dragged to any other point of $\CB^*$ by following a path $\g$ connecting $p$ to $p^\g$,
\begin{align}
    \d(p) \overset\g\rightsquigarrow \d(p^\g) . 
\end{align}
Since $\CB^*$ is not simply connected, the basis can suffer monodromies by $\Sp(4,\Z)$ transformations (since the intersection pairing is preserved under monodromy),
\begin{align}\label{1-cycle monod}
    M: \pi_1(\CB^*) &\to \Sp(4,\Z) ,&
    \d_a & \overset\g\rightsquigarrow (\d_a)^\g = M(\g)^b_a \, \d_b,
\end{align}
where $\g\in\pi_1(\CB^*)$ is a closed path on $\CB^*$.  
We will often neglect to write the dependence on $\g$ when it is understood from context.
Thus, in general there may be no globally defined choice of 1-cycle basis on CB$^*$. As we will soon show, in the presence of a 1-form basis monodromy, there is a slight difference between $M$ and the standard EM duality monodromy group of SW solutions. 

Taking this into account, points of $\cH_2$ correspond to equivalence classes of RSs with ``marking" --- a choice of canonical 1-cycle homology basis.
The $\Sp(4,\Z)$ group of 1-cycle basis changes acts on $\t$ by the generalized M\"obius transformation $M\circ\t$, given in \eqref{mobius}.
Two points, $\t, \t' \in \cH_2$, describe the same (possibly split) genus-2 RS iff they are related by an $\Sp(4,\Z)$ transformation, $\t' = M \circ \t$, $M\in \Sp(4,\Z)$.
A fundamental domain, $\cF_2$, of the $\Sp(4,\Z)$ action on $\cH_2$ is described in appendix \ref{appZ}.
Thus $\cF_2$, together with appropriate identifications of its boundaries, is the moduli space of (possibly split) genus-2 RSs, i.e., $\cF_2 \simeq \cH_2/\Sp(4,\Z)$.

Since $\t_{ij}$ up to choice of 1-cycle basis is a physical observable, it follows that the SW curve as family of abstract RSs is holomorphic (i.e., single-valued) on CB$^*$: there are no monodromies in the complex structure of $\S$ over $\CB^*$.
This does not contradict the existence of 1-cycle basis monodromies \eqref{1-cycle monod} since those correspond to monodromies in a choice of marking on the RS, and not to monodromies of its complex structure.

\subsubsection{The 1-form basis monodromy}

We now will show that the single-valuedness of the low energy physics on the CB allows for monodromies also in the 1-form basis. Both the special coordinates \eqref{specialcoords} and the Kahler metric on the CB, which is given by the Kahler potential
\begin{align}\label{KahlerK}
    K = \frac i2 J(\s,\bar\s) = \Im (s^D_i \bar s^i) ,
\end{align}
depend on the choice of 1-form basis and will be affected by these monodromies.

To see this, consider the periods $\Pi_a^j(p)$ in \eqref{periodmat}. Since they are just derivatives of the special and dual special coordinates, which along closed loops vary by a $(u,v)$ independent $\Sp(4,\Z)$ EM monodromy group element, there must exist an EM monodromy matrix $\til M^b_a \in \Sp(4,\Z)$ such that
\begin{align}\label{Pi monod0}
    \Pi^j_a(p) \overset\g\rightsquigarrow \Pi^j_a(p^\g) 
    = \til M(\g)_a^b \, \Pi_b^j(p) ,
\end{align} 
for each $\g\in\pi_1(\CB^*)$. This is consistent with physical single-valuedness of the central charge \eqref{delZ} and metric \eqref{KahlerK}. The crucial point here is that in general $M(\g)\neq \widetilde{M}(\g)$ and the mismatch between the two is accounted for precisely by an extra monodromy in the 1-form basis.

Explicitly, say that the 1-form basis suffers the monodromy\footnote{The argument below will show that only the conjugacy class of $F$ in $\GL(2,\C)$ is homotopy invariant, while the particular representative of that conjugacy class may vary with the base point; hence the dependence on the choice of base point $p\in\CB^*$ in \eqref{basis monod}.}
\begin{align}\label{basis monod}
    F: \pi_1(\CB^*, p) &\to \GL(2,\C) , &
    \Om_j(p) \overset\g\rightsquigarrow \Om_j(p^\g) = 
    \Om_k(p) F(\g,p)_j^k .
\end{align}
From the definition of the period matrix, upon traversing $\g\in\pi_1(\CB^*,p)$, it suffers monodromy
\begin{align}\label{Pi monod}
    \Pi(p) \overset\g\rightsquigarrow \Pi(p^\g) 
    = M(\g) \,\, \Pi(p) \, F(\g,p),
\end{align} 
where $M(\g)$ is the 1-cycle basis monodromy associated to $\g$ and we dropped the $a,b=1,\ldots,4$ and $i,j=1,2$ indices, working instead in an obvious matrix notation.
Comparing to \eqref{Pi monod0} gives
\begin{align}\label{period monod}
    M \, \Pi \, F = \til M \, \Pi
\end{align}
everywhere on CB$^*$ and for $M$ and $\til M$ depending on each $\g\in\pi_1(\CB^*)$.
We have dropped the explicit $\g$ and base point dependencies, leaving them as understood from now on.

To derive the constraints on the possible 1-form basis monodromies $F$ from \eqref{period monod}, rewrite \eqref{period monod} as $\Pi F = N \Pi$ with $2\times2$ block decompositions
\begin{align}\label{Ndef}
    \bpm A & B \\ C & D \epm &= N \doteq M^{-1} \til M \in \Sp(4,\Z) ,&
    &\text{and} &
    \Pi &= \bpm \t \\ 1 \epm \cA .
\end{align}
This readily implies that $N$ fixes $\t$, \emph{i.e.}
\beq\label{FandN1}
N\circ \t=\t
\eeq
where $N\circ \t$ denotes the M\"obius action of $\Sp(4,\Z)$ on $\cH_2$ given in \eqref{mobius}, and also implies that 
\begin{align}\label{FandN}
    F &= \cA^{-1} (C\t+D) \, \cA .
\end{align}

The set of $N$ which fix a given $\t\in\cH_2$ form a finite subgroup of $\Sp(4,\Z)$ which is the image of the automorphism group of the RS with complex structure $\t$, as we will review in the next section.
The automorphism group, $\aut(\t)$, of a RS with modulus $\t$ is the group of its isomorphisms;
it is a finite group for genus 2 or greater.
Its action on the points of the RS induces an $\Sp(4,\Z)$ action on a choice of 1-cycle basis, and also induces a $\GL(2,\C)$ action on a choice of 1-form basis, which we will denote, respectively, by the morphisms
\begin{align}\label{aut morphisms}
    N &: \aut(\t) \to \Sp(4,\Z), &
    F &: \aut(\t) \to \GL(2,\C).
\end{align}
They are related by \eqref{FandN1} and \eqref{FandN} and it is straightforward to show that $N$ is trivial iff $F$ is.

We review the definition of automorphism groups of genus 2 RSs presented as binary sextic plane curves in the next section. 
In appendix \ref{Autsec} we show how to explicitly construct their associated representations \eqref{aut morphisms}.

\subsection{The automorphism twist}

The above discussion implies that the physically allowed 1-form basis monodromies are determined in terms of a representation of the fundamental group of $\CB^*$ in the automorphism group, $\aut$, of the SW curve (the holomorphic family of RSs over CB$^*$),
\begin{align}\label{aut twist}
    \a : \pi_1(\CB^*) \to \aut .
\end{align}
In fact to each closed path $\g\in\pi_1(\CB^*)$ starting and ending at a base point $p\in\CB^*$, there is associated a 1-form basis monodromy $F(\g,p)$.  
This 1-form basis monodromy is determined by \eqref{aut twist} via constructing the associated $\Sp(4,\Z)$ element $N(\a(\g))$, and then constructing $F(\a(\g)) \in \GL(2,\C)$ using \eqref{FandN} and the ``electric" period matrix $\cA$.
Note that $\cA$ depends not only on the RS, but also on the choice of 1-cycle basis, as well as on the 1-form basis.
As such, it also depends on the base point $p\in \CB^*$. We will call the monodromy \eqref{aut twist} the \emph{automorphism twist} of the SW data.

The automorphism twist is a new entry we are introducing to describe the SW data specifying a rank-2 CB geometry.
Given a holomorphic family of RSs over CB$^*$ with an associated integrable 1-form basis on the CB, then the automorphism twist of this family is unambiguously determined.
In particular, the twist is non-trivial if and only if the 1-form basis monodromy is non-trivial.
So, allowing 1-form bases to be multi-valued on CB$^*$ requires the introduction of an automorphism twist.

But the automorphism twist is \emph{not} an invariant of the SK structure of the CB.
A given SK structure can be described by many different sets of holomorphic families of RSs and automorphism twists, $(\text{RS},\a ) \simeq (\text{RS}',\a')$.
Furthermore, we show in the next section that each SK structure can be described by a holomorphic family of RSs with a \emph{periodic} family of integrable 1-form bases, and so with trivial automorphism twist.

So, if we do not need them, why consider automorphism twists and non-periodic 1-form bases at all?
The reason is calculational convenience.
It turns out to be algebraically quite complicated to solve for all holomorphic families of genus 2 RSs with periodic integrable 1-form bases.
For instance, in \cite{Argyres:2005pp} such an attempt even with additional (and not physically well-motivated) simplifying assumptions resulted in a system of 33 5th order polynomial equations in 21 unknowns. 
We will show below, by way of examples, that for families of genus 2 RSs with enhanced automorphism groups solutions can be found more easily by restricting the algebraic form of the family of RSs but allowing automorphism twists.

Before doing so, it will be useful to clarify the relation of the automorphism twist to the behavior of the family of RSs, and to physical observables.
We turn to this now.

\subsubsection{Geometrical interpretation of the automorphism twist, and the EM duality monodromy}

Consider invariants of the SK structure on the CB.
One obvious invariant is the matrix of electromagnetic couplings, $\t$, thought of as a holomorphic function
\begin{align}
    \tau : \CB^* \to \cH_2/\Sp(4,\Z) ,
\end{align}
into the moduli space of (possibly split) genus-2 RSs.
As remarked earlier, this invariant does not depend on the 1-form basis, and therefore is insensitive to the automorphism twist.

A more interesting invariant, which is sensitive to the automorphism twist, is the \emph{EM duality monodromy} of the CB,
\begin{align}\label{duality monod}
    \til M &: \pi_1(\CB^*) \to \Sp(4,\Z) ,
\end{align}
the $\Sp(4,\Z)$ monodromy of the special coordinates, and thus of the period matrix \eqref{Pi monod0}.
This is the monodromy of the low energy theory on the CB which is physically observable.
It need not coincide with either the 1-cycle basis or with the automorphism twist monodromy, but instead is, in general, a combination of both.
Indeed, by \eqref{Ndef}, the EM duality monodromy is given by
\begin{align}\label{duality monod 2}
    \til M(\g) = M(\g) N(\a(\g)) ,
\end{align}
where $M(\g)$ is the 1-cycle basis monodromy and $N(\a)$ is the $\Sp(4,\Z)$ representation of the automorphism twist, associated to the 1-form basis monodromy.
We interpret this as saying that the holomorphic family of RSs has 1-cycle basis monodromy $M$ plus an additional basis ``jump" by $N(\a)$ induced by the automorphism twist.

To see this, we work on the \emph{cut CB}, where $\CB_{cut}$ is $\CB^*$ minus a cut $C$ across which the multivaluedness is ``concentrated''. This allows a clear description of the multivaluedness of the 1-form basis as boundary conditions across the cuts in a two complex dimensional analogue of familiar branch cuts of multi-valued functions on RSs.\footnote{More formally, the ``cut" $C = \coprod_i C_i$ is a union of disjoint 3-real-dimensional manifolds $C_i$ whose boundaries are components of $\cV$ and such that the cut CB is a simply connected complex manifold. Or, equivalently, $\CB_{cut}$ is the interior of a connected fundamental domain of the action of $\pi_1(\CB^*)$ on the universal cover of $\CB^*$.
The boundary of the cut CB is thus $\del\CB_{cut} = \coprod_i (C_i \coprod C_i^{\g_i})$ which are pairwise identified by a set of homotopy closed paths generating the fundamental group, $\vev{\g^i} = \pi_1(\CB^*)$.
If $p\in C$ and $\g\in\pi_1(\CB^*)$, then the lift of $\g$ to $\CB_{cut}$ begins and ends at two boundary points $p\in C_i$ and $p^\g\in C_i^\g$.}

For $p\in\CB^*$, denote the RS at that point by $\S(p)$.
Analytically continue the RS at some point $p\in C$ in $\del\CB_{cut}$ along a closed path $\g \subset \CB^*$, which ends at $p^\g$ also lying on  $ \del\CB_{cut}$.
The analytic continuation
\begin{align}\label{Sac}
    \S(p) \overset\g\rightsquigarrow \S(p^\g) ,
\end{align}
depends only on the homotopy class of $\g \in \pi_1(\CB^*)$.
Since the family of RSs is single-valued over $\CB^*$, $\S(p^\g)$ must describe the same RS as $\S(p)$, i.e., there must be some automorphism $\a(\g) \in \aut(\S)$ such that
\begin{align}\label{SBC}
    \S(p^\g) = \a(\g) \circ \S(p) ,
\end{align}
where $\circ$ denotes the action of the automorphism on the RS.
This automorphism twist is not a property of an abstract family of RSs since, by definition, automorphisms leave the RS unchanged.
But, upon choosing a cut, the multi-valuedness \eqref{basis monod} of the 1-form basis is described as a ``jump" across that cut, and it is natural to localize the associated automorphism action at that cut.
Thus we arrive at the following picture: a SW data in the presence of an automorphism twist is a holomorphic family of RSs and 1-form bases varying over the cut CB with the additional action of the automorphism twist specifying how the family is glued together across the cut.
We refer to these gluing conditions, \eqref{basis monod} and \eqref{SBC}, as the automorphism twist boundary conditions on the cut CB.

For example, in case $\a(\g) =1$, the identity automorphism, then $N(\a(\g)) = 1 \in \Sp(4,\Z)$, and, by \eqref{FandN}, $F(\g,p) = 1 \in \GL(2,\C)$.
Thus when there is no automorphism twist, the 1-form basis is single-valued (holomorphic on $\CB^*$), and the EM duality monodromy coincides with the 1-cycle basis monodromy.
This is the case examined in \cite{Argyres:2022lah, AM22II, Argyres:2005pp, Argyres:2005wx}.
We show in the next section that any SW data on the CB can be brought to this form with trivial automorphism twist by suitable coordinate redefinitions.

An example in the opposite extreme situation is one in which the family of RSs is globally constant over $\CB^*$, i.e., in which the total space of the fibration of the family of RSs over $\CB$ is the direct product $\text{RS}(\t) \times \CB$.
Then the 1-cycle basis monodromy is trivial, and the EM duality monodromy coincides with the automorphism twist monodromy.
This situation may seem very special, since it could only occur if the coupling matrix $\t$ were constant over the CB.
But, in fact, it turns out that many $\cN=2$ SCFTs have this property.
For example, all $\cN{=}4$ sYM theories, all $\cN{=}3$ SCFTs \cite{Seiberg:1994bz,Cordova:2016xhm,Argyres:2019yyb}, and all $\cN{=}2$ SCFTs with characteristic dimension not equal to 1 or 2 \cite{Cecotti:2021ouq}, have constant $\t$.

\section{Rank-2 CBs described by binary-sextic plane curves}
\label{secBS}

This section is devoted to translating the monodromy conditions on the RS and 1-form basis to the $a_i^j$ and $c_n$ coefficient functions in \eqref{SWcrv}-\eqref{1fbasis2} that enter into our chosen algebraic presentation --- i.e., as a binary sextic plane curve with holomorphically varying 1-form basis. 

We start by discussing the coordinate reparameterization freedom of a binary sextic curve plus 1-form basis, and its relation to automorphisms of genus 2 RSs.
We then characterize what multivaluedness of the curve and 1-form basis is compatible with the physical monodromy conditions derived in the previous section.

\subsection{$\GL(2,\C)/\Z_3$ reparametrizations and automorphisms}

Coordinate transformations of the form
\begin{align}\label{Gmap}
x &\mapsto \frac{G^1_1 x+ G^2_1}{G^1_2 x+G^2_2} , &
y &\mapsto \frac y{(G^1_2 x+ G^2_2)^3} ,&
\bspm G^1_1 & G^2_1 \\ G^1_2 & G^2_2 \espm &\in \GL(2,\C),
\end{align}
preserve the binary-sextic form of the curve, though not the specific values of the $c_n$ coefficients of the curve.  
In fact, they map the coefficients as $c_n \mapsto c'_n := G\circ c_n$ where $G\circ c_n$ depend linearly on the original $c_n$, and in a complicated way (but homogeneously of degree 6) on the $G^i_j$. In addition to \eqref{Gmap}, there is a freedom to rescale $(u,v)$ by an overall complex normalization
which enhances to a $\GL(2,\C)$ reparametrization invariance when $u$ and $v$ have the same scaling dimension. This freedom was called \emph{holomorphic reparametrization invariance} in \cite{Argyres:2005pp} and is used in \cite{Argyres:2022lah} to set non-vanishing parameters of the SW curve to some fixed values. We neglect this here and henceforth only discuss the implications of \eqref{Gmap}.

It is useful to note that this transformation maps the discriminant of $c(x,u,v)$ as
\begin{align}\label{Gdmap}
    G : \text{Disc}_x(c) \mapsto (\det G)^{30} \,\text{Disc}_x(c) .
\end{align}  
Here the exponent $30=d(d-1)$ where $d$ is the degree of the polynomial $c$ in $x$.%
\footnote{In case one of the roots of $c(x)$ is at infinity so that the $y^2=c(x)$ curve is a binary quintic, then if $G^1_2=0$, $G\circ$ maps the curve to another binary quintic, and the discriminant maps to $(\det G)^{20} \,\text{Disc}_x(c)$.} 
This shows, in particular, that the discriminant of the curve is not invariant under reparameterizations.

The $\GL(2,\C)$ transformation acts on the canonical 1-form basis \eqref{1fbasis1} as
\begin{align}\label{Gwmap}
G : \w_i &\mapsto (\det G)\, \w_j \, G_i^j  ,
\end{align}
and thus on the scaling coordinate 1-form basis as
\begin{align}\label{Gamap}
G : \Om_i &\mapsto \til\Om_i := \w_j \, \til a_i^j , & 
&\text{with} &
\til a &:= (\det G) \, G \, a  .
\end{align}
So it maps the basis of one forms by a proportional $\GL(2,\C)$ element, $(\det G)\, G$, whose determinant is thus $(\det G)^3$.  

When the transformation is in the $\Z_3$ in the center of $\GL(2,\C)$ consisting of elements of the form
\begin{align}\label{Z3center}
    \Z_3 &:= \left\{ \z_3^n I_2 , \ n=0,1,2 \right\}, &
    \z_3 &:= e^{2\pi i/3}, &
    I_2 &:= \bspm1&0\\0&1\espm,
\end{align}
it acts trivially on the points of the curve and the 1-forms.
So the nontrivial coordinate transformations are those in $\GL(2,\C)/\Z_3$.

\subsubsection{Automorphism jumps}

An automorphism of a genus-2 RS presented as a binary sextic plane curve is a $\GL(2,\C)/\Z_3$ reparameterization which leaves the curve describing the RS unchanged, \emph{i.e.} $c_n=G \circ c_n$ iff $G$ implements the automorphism.  

An automorphism of any genus-2 RS is $G = [- I_2] :=-1$, which acts on the RS coordinates and 1-forms as
\begin{align}\label{G-1}
    -1:&& (x,y) &\mapsto (x,-y),& \w_i &\mapsto -\w_i ,
\end{align}
and so obviously leaves the $y^2=c(x,u,v)$ plane curve invariant.
This is the hyperelliptic automorphism that all genus-2 RSs enjoy, and it is the only non-trivial automorphism that a generic genus-2 RS has.
But there are special subvarieties of the moduli space of genus-2 RSs whose members have ``extra" automorphisms.
These larger automorphism groups are all finite, and are listed in appendix \ref{Autsec}. 

Let $\aut(\t)$ be an isomorphism of RS$(\t)$, where we take $\t\in\cF_2$.
$\aut(\t)$ acts trivially on $\cF_2$ by definition, but acts non-trivially on the points of RS$(\t)$, via reparameterizations,
\begin{align}\label{ad1}
    G_\aut: \aut(\t) \to \GL(2,\C)/\Z_3 .
\end{align}
This reparameterization map induces the $N$ map of 1-cycles bases \eqref{aut morphisms}.

The $G_\aut$ and $N$ representations of $\aut(\t)$ are related as follows.
On the canonical 1-form basis, $\w$, thought of as a 2-component row vector, $G_\aut$ acts by right matrix multiplication,
\beq
G_\aut: \w \mapsto (\det G_\aut) \, \w \, G_\aut.
\eeq
For a given choice of 1-cycle basis, $\d$, thought of as a 4-component column vector, $N$ acts by left matrix multiplication, $N: \d \mapsto N\d$.  
Then the $4\times 2$ canonical period matrix, $\Pi_\w \doteq \oint_\d \w$, transforms as 
\beq
\Pi_\w \mapsto (\det G_\aut)\, N\, \Pi_\w\, G_\aut.
\eeq
But since this is an isomorphism these period matrices are equal. Comparing the equation above with \eqref{period monod} and recalling that $N\doteq M^{-1}\widetilde{M}$ and that $\Om\overset\g\rightsquigarrow \Om F$ where $\Om=\w\, a(p)$ is a general holomorphic 1-forms basis, 
we readily find that $F$ is conjugate to the inverse of the reparameterization representation, $G_\aut$, of the automorphism group:
\beq\label{FGaut}
F(\g,p)=a^{-1}(p)(\det G_\aut)^{-1}\, G_\aut^{-1}\, a(p).
\eeq 
\eqref{FGaut} is an important equation which will be used below to derive the consistent boundary conditions of the SW data.

Now, write $\Pi_\w = \bspm \t\\1\espm \cA_\w$ as in \eqref{Pitotau}, where we are keeping track of the fact that the periods are in the canonical 1-form bases $\w$, $\t$ is the modulus of the RS, and $\cA_w$ is the $2\times2$ ``electric" canonical period matrix in our given 1-cycle basis, see \eqref{specialcoords}.
Then
\begin{align}\label{ad3}
    (\det G_\aut)^{-1} \, G_\aut^{-1} &=
    \cA_\w^{-1} (C\t+D) \cA_\w, &
    N \circ \t &= \t, &
    N &\doteq \bspm A&B\\ C&D\espm.
\end{align}
In summary, for a given RS specified by $\t \in \cF_2$, $N(\aut(\t))$ is determined to be the finite subgroup of $\Sp(4,\Z)$ fixing $\t$.
Then, given a specific binary-sextic curve realizing RS$(\t)$, and a choice of 1-cycle basis, the period matrix $\cA_\w$ is determined, and \eqref{ad3} uniquely determines $G_\aut \in \GL(2,\C)/\Z_3$ in terms of $N$. 

Finally consider a family of RSs varying holomorphically over CB$^*$.
If the RSs in a neighborhood of any point of CB$^*$ have a non-trivial automorphism, then the whole family must have that automorphism, by analytic continuation.
Thus the automorphism group is a property of the whole family, and we can speak of the automorphism group of the SW curve.%
\footnote{It is possible that RSs over complex subvarieties of CB$^*$ have enhanced automorphism groups, so perhaps it would be more accurate to speak of the generic automorphism group of the SW curve.}

\subsubsection{Including reparametrizations}

The previous subsection derived an explicit expression of the 1-form basis monodromy in terms of the $GL(2,\Z)/\Z_3$ matrix implementing the automorphism and the 1-form basis matrix $a(p)$. 
We now show how the 1-form basis monodromy is related to monodromies of the SW curve coefficients.

This is unfortunately not the end of the story. In fact the automorphism group is by definition the subgroup of reparametrization transformations on the $(x,y)$ which leaves each coefficient of the curve invariant. 

Upon analytically continuing the SW curve from some point $p\in\CB^*$ along a closed path $\g \subset \CB^*$, the RS does not change, but the coefficients of the SW curve describing the RS can change by an arbitrary $\GL(2,\C)/\Z_3$ coordinate reparametrization,
\begin{align}\label{cBC}
    c_n(p) \overset\g\rightsquigarrow c_n(p^\g) = H(\g,p) \circ c_n(p) ,
\end{align}
where $H(\g,p) \in \GL(2,\C)/\Z_3$ and $\circ$ denotes the $\GL(2,\C)/\Z_3$ action on the $c_n$ \eqref{Gmap}.
In other words, the $c_n$ are locally holomorphic functions which can be branched over the singularity locus $\cV \subset\CB$. 
Note that $H(\g,p)$ can vary holomorphically with respect to the base point $p$.

The periodicities \eqref{cBC} also affects the expression for the 1-form basis, which explicitly depend on $(x,y)$, and need to be ``added'' to the automorphism jump discussed in the previous section. 
This analysis closely parallels the one in the previous subsection.
Recall the form of the holomorphic 1-form basis, $\Om$, on the RS, $\Om = \w \, a$. 
The $2\times2$ 1-form basis coefficient matrix $a$ analytically continues to $a(p) \overset\g\rightsquigarrow a(p^\g)$. 
Similarly the canonical 1-form basis $\w$ transforms as \eqref{Gwmap} under the extra $\GL(2,\C)/\Z_3$ reparameterization $H(\g,p)$ \eqref{cBC}. 
Bringing everything together we have 
\begin{align}\label{1fmonod}
    \Om(p)
    \overset\g\rightsquigarrow
    \Om(p^\g) 
    &=
    \w(p^\g) a(p^\g) 
    = 
    \det[H(\g,p)] \, \w(p) \, H(\g,p) \, a(p^\g)  \nn\\
    &= 
    \det[H(\g,p)] \, \Om(p) \, a^{-1}(p) \, H(\g,p) \, a(p^\g) .
\end{align}
On the other hand, this has to be compatible with the consistency conditions following from single-valuedness of the low energy physics on the CB,
\beq
\Om(p^\g) = \Om(p) \, F(\g,p) ,
\eeq
where $F$ satisfies \eqref{FandN} and accounts for the monodromy jump that only comes from the automorphism part of the reparametrization transformation, $G_\aut$.
Recall here that $G_\aut$ leaves the $c_n$ invariant, $c_n=G_\aut \circ c_n$, while $H(\g,p)$ does not.
Comparing this with \eqref{1fmonod}, we find the monodromy condition on the 1-form basis matrix
\beq
a(p^\g) =  \det[H(\g,p)]^{-1} \,H^{-1}(\g,p) \, a(p) \, F(\g,p),
\eeq
from which, using \eqref{FGaut}, we arrive at the most general consistent monodromy of 1-form bases,
\begin{align}\label{aBC}
    a(p^\g) &=
    \det[G_\aut H]^{-1} \, (G_\aut H)^{-1} \, a(p) ,
\end{align}
where we have suppressed the $(\g,p)$ dependence of $H$ and $G_\aut$.
$G_\aut$  inherits its path-dependence from the automorphism twist: $G_\aut(\g,p) = G_\aut(\a(\g))(p)$.

Before continuing our presentation, let's pause briefly to summarize what we have done. As we move around closed loops on $\CB^*$, the allowed change in the SW data receives contributions from two sources, an automorphism jump and a general reparametrization transformation. 
The former does not act, by definition, on the coefficients of the curve but changes the 1-form basis as in \eqref{FGaut}. 
The latter modifies both the coefficients of the curve and the 1-form bases. 
Equations \eqref{cBC} and \eqref{aBC} bring both contributions together to give the periodicity conditions the binary sextic curve $y^2=c(x,u,v)$ and 1-form basis $\Om = \w a$ coefficient functions must satisfy on CB$^*$, including a given automorphism twist. 
The freedom to change both the curve coefficients and the 1-form basis by $\GL(2,\C)/\Z_3$ reparameterizations makes the periodicity conditions \eqref{cBC} and \eqref{aBC} difficult to implement. 
The solution is to work in ``coordinate frames'' in which the $\GL(2,\C)/\Z_3$ invariance is fixed by imposing suitable conditions on 1-form basis and the curve. 
We will discuss this next.

\subsection{The canonical reparametrization frames}

Choose a reparameterization which varies over the CB in such a way that the 1-form basis is the canonical basis everywhere on CB$^*$,
\begin{align}\label{cfa}
\Om_i = \w_i .
\end{align}
This is possible since $a$ is holomorphic and invertible on $\CB^*$.
In particular, act with the $\GL(2,\C)/\Z_3$ transformation given by $M= a^{-1} (\det a)^{1/3}$ on the cut CB.
Since $\det a \neq 0$ on the cut CB (it can only vanish on the singular loci which are in the boundaries of the cut CB), a global cube root of $\det a$ can be chosen.
(The phase of the cube root is trivial in $\GL(2,\C)/\Z_3$.)
Then by \eqref{Gamap}, the transformed 1-form basis matrix is $\til a=1$.

We call this the \emph{canonical frame}; choosing it completely fixes the reparameterization freedom.
The canonical frame ``flattens" the 1-form basis coefficients so that $a=1$ everywhere on the cut CB.
In particular, in this frame there is no 1-form basis monodromy, so the automorphism twist is trivial, $G_\aut(\g,p) = 1$.  
Since the 1-form basis coefficients are periodic, $a(p^\g) = a(p) = 1$, \eqref{aBC} and the triviality of $G_\aut$ imply also that $H(\g,p)=1$ for all $p$ and $\g$.
Thus, by \eqref{cBC}, the curve coefficients are single-valued on $\CB^*$.
Since they are locally holomorphic on CB$^*$, this means they are holomorphic on CB$^*$.
Furthermore, because of the regularity conditions that physical CB geometries must satisfy on the singular locus $\cV\subset\CB$ \cite{Argyres:2018zay, Argyres:2018urp, Argyres:2020wmq}, this means that the curves are in fact meromorphic on the whole CB.

In summary, in the canonical frame the reparametrization freedom is completely fixed, the 1-form basis is the canonical basis, and the binary-sextic curve is meromorphic on the CB.\footnote{Here we are not considering the holomorphic reparametrization invariance of the CB itself, which is not fixed by the choice of canonical frame and can be used to further set some coefficients of $c(x,u,v)$ to convenient values.}
Furthermore, any SW data can be brought to canonical frame by a suitable reparameterization.

In the canonical frame the SK integrability conditions \eqref{SKint1} simplify to the system of 8 ordinary differential equations for 11 unknown functions described and analyzed in \cite{Argyres:2005pp, Argyres:2005wx, Argyres:2022lah, AM22II}.
Unfortunately, as described in those references, the task of finding all the solutions is algebraically daunting.
So far only an incomplete list of solutions has been found by imposing simplifying assumptions.

We will now propose a set of coordinate frames which are different from the canonical frame, but in which the SK integrability conditions might become more tractable.
We call these frames \emph{automorphism frames} because they are closely tied to the automorphism group of the family of curves.

\subsection{Automorphism frames}
\label{aut frames}

Suppose the curves of a family of RSs, $\S(s)$, varying holomorphically with some parameters, $s$, and with a given automorphism group, aut, can be put by suitable $\GL(2,\C)/\Z_3$ transformations into a canonical parameterization,
\begin{align}\label{autcrv}
    y^2 = c^\aut(x) := \sum_{n=0}^6 c_n^\aut(s) x^n ,
\end{align}
such that $c^\aut_n(s)=c^\aut_n(s')$ iff $\S(s)=\S(s')$ as RSs.
We discuss below when and how this can be done.
We will call a choice of $c^\aut(x)$ fixing the reparameterization freedom an ``automorphism frame".  
This should be contrasted with the canonical frame introduced in the previous subsection.
The canonical frame fixes the reparameterization freedom by choosing a canonical basis for the holomorphic 1-forms, while the automorphism frame fixes it by choosing a canonical form for the curve.

In an automorphism frame the only unfixed part of the reparameterization group are the automorphisms of the curve.
Since, by definition, these do not change the coefficients of the curve, any holomorphic family of RSs with this automorphism group over CB$^*$ will have automorphism frame curves whose coefficients are also holomorphic (i.e., periodic) on CB$^*$, and hence meromorphic on CB.
Thus knowledge of the automorphism symmetry is leveraged to fix some (or all) curve coefficients.
Clearly the bigger the automorphism group, the tighter the constraints.

The calculational price that is paid for this simplification is that the 1-form basis is no longer constant in this frame, and so its coefficient functions enter into the SK integrability differential equations.
Furthermore, the 1-form basis coefficients are no longer periodic on $\CB^*$ but suffer automorphism twist monodromies.
But, at least for large enough automorphism groups, the resulting integrability differential equations simplify dramatically to a system of \emph{linear} ordinary differential equations, and the monodromy group of these differential equations is closely tied to the automorphism twist monodromy representation.
This can make solving for the 1-form basis very easy, as we show in an example in section \ref{sec4}.

There is no unique choice for the automorphism frame form of the curve.
It should be chosen in such a way as to simplify the form of the integrability equations.
We have made some suggestions of simple-looking forms of $c^\aut(x)$ in table \ref{tabA} --- see appendix \ref{Autsec} for a detailed explanation of the table.

In the five cases where there is either a unique RS with given automorphism group (i.e., when $\aut=\GL_2(3)$, $\Z_{10}$, or $\Z_2\rtimes D_4$) or a 1-parameter family of RSs (i.e., when $\aut=D_4$ or $D_6$), then the automorphism frames given in table \ref{tabA} do the job of completely fixing the reparameterization freedom down to just the automorphism group.
In the remaining two cases where there is a 2-parameter family of RSs with $\aut=D_2$ and the generic (3-parameter) family with just the hyperelliptic automorphism $\aut=\Z_2$, the automorphism frame curves listed unfortunately do not completely fix the reparameterization freedom down to the automorphism group.

For instance, in the $D_2$ case with $c^\aut(x)$ chosen to be of Legendre type, $y^2 = (x^2-1) (x^4 + c_2 x^2 + c_0)$, a calculation shows that a $\Z_4$ subgroup of reparameterizations generated by $\GL(2,\C)/\Z_3$ transformations which map $x\mapsto 1/x$ and $x\mapsto \sqrt{r_+}\, x$ where $r_\pm \doteq (-c_2\pm\sqrt{c_2^2-4c_0})/2$, act within this family of curves by mapping their coefficients as
\begin{align}\label{D2map1}
    x &\mapsto \frac1x , &
    (c_0,c_2) &\to \Bigl(\, \frac1{c_0} \,,\,\frac{c_2}{c_0} \, \Bigr) ,\\
    x &\mapsto \textstyle{\sqrt{r_+}} \, x , &
    (c_0,c_2) &\to \Bigl( \frac{c_2}{c_0}-r_- \frac{c_0-c_2^2}{c_0^2} 
    \ ,\ 1+ r_+ \frac{c_2-1}{c_0} \, \Bigr) . \nn
\end{align}
Systematically incorporating the possibility of these complicated periodicities in the $c_n$ coefficients on CB$^*$ seems difficult.

\begin{table}[t!]
\centering
$\begin{array}{c|c|c|c|c|c}
\aut & \bar\aut & c^\aut(x) & \bar\aut\subset\PSL(2,\C) & \t_{ij} 
& \text{\footnotesize dim$\cM$} \\ 
\hline
\Z_2 & 1 & x(x{-}1) (x^3 {+} c_2 x^2 {+} c_1 x {+} c_0)
& \langle x{\mapsto} x\rangle 
& \bspm z_1 & z_2\\ z_2 & z_3 \espm & 3\\[2mm]
D_2 & \Z_2 & (x^2{-}1)(x^4 {+} c_2 x^2 {+} c_0)
& \langle x{\mapsto} {-}x\rangle 
& \bspm z_1 & z_2\\ z_2 & z_1 \espm & 2\\[2mm]
D_4 & D_2 & (x^2{-}1)(x^4 {+} c_2 x^2 {+} 1)
& \big\langle x{\mapsto}{-}x , x{\mapsto}\frac1x \big\rangle 
& \frac12 \! \bspm z & 1\\ 1 & z \espm &1\\[2mm]
D_6 & S_3 & x^6 {+} c_3 x^3 {+} 1
& \big\langle x{\mapsto} \z_3 x , x{\mapsto}\frac1x \big\rangle 
& z\bspm 2 & 1\\ 1 & 2 \espm &1\\[2mm]
\GL_2(3) & S_4 & x(x^4{-}1)
& \big\langle x{\mapsto} \z_4 x, x{\mapsto}\frac1x, 
x{\mapsto}\frac{x-1}{x+1}\big\rangle 
& \frac{i\sqrt2 -1}3 \! \bspm 2 & 1\\ 1 & 2 \espm &0\\[2mm]
\Z_{10} & \Z_5 & x(x^5{-}1)
& \langle x{\mapsto} \z_5 x \rangle 
& \bspm \z_5 & \z_5+\z_5^3\\ \z_5+\z_5^3 & -\z_5^4 \espm &0\\[2mm]
\Z_3{\rtimes} D_4 & D_6 & x^6{-}1
& \big\langle x{\mapsto} \z_6 x , x{\mapsto}\frac1x \big\rangle 
& \frac{i}{\sqrt3} \! \bspm 2 & 1\\ 1 & 2 \espm &0
\end{array}$
\caption{\label{tabA} 
Some data describing the automorphism groups of genus 2 RSs.}
\end{table}

A better choice of automorphism frame in this case may be to impose instead a Weierstrass-type form for the curve, $y^2 = x^6 + c_4 x^4 + c_2 x^2 + c_0$, together with the constraint on the 1-form basis that $(\det a)^2=1$.
A calculation shows that the unfixed reparameterization group modulo the $D_2$ automorphism group is the $\Z_3$ generated by
\begin{align}\label{D2map5}
    (x,y) &\mapsto (\z_3^2 x, y), &
    (c_0,c_2,c_4) &\to \bigl(\, c_0, \z_3 c_2, \z_3^2 c_4 \bigr) ,
\end{align}
as long as $c_0 \neq \pm1$.  
These possible cube root phase periodicities should be easier to systematically incorporate.

Similarly difficult is the generic case where $\aut=\Z_2$.
For example, for the ``Legendre-like" form for $c^\aut(x)$ shown in the first row of table \ref{tabA} there is an $S_5$ subgroup of $\GL(2,\C)/\Z_3$ transformations which keep $c^\aut(x)$ degree five, monic, and with roots at $0$ and $1$, but which redefine the $c_n$.  
Most of these give rise to non-rational transformations of the coefficients of $c^\aut(x)$ which seem unlikely to give rise to consistent boundary conditions on the cut CB.
But even so, there is an $S_3$ subgroup of rational transformations of the curve coefficients, analogous to the $S_3$ group of rational identifications of $\l$ in the Legendre form of an elliptic curve.
The existence of this $S_3$ of additional rational identifications of curve coefficients makes working in an automorphism frame with the Legendre-like form of $c^\aut(x)$ difficult.

A better gauge fixing is to work instead with a Weierstrass-like form of the curve, $y^2 = x^5 + c_3 x^3 + c_2 x^2 + c_1 x + c_0$, together with the constraint on the 1-form basis that $\det a=1$.
Then the unfixed subgroup of $\GL(2,\C)/\Z_3$ which acts within this frame is the $\Z_4$ generated by
\begin{align}\label{hyperframe}
    x &\mapsto -x, & y &\mapsto i y, &
    c_n &\mapsto (-)^{n+1} c_n .
\end{align}
(The $\Z_2 \subset \Z_4$ are the hyperelliptic automorphisms.)
Again, these square-root phase periodicities of the curve look likely to be amenable to systematic analysis.

We leave to future work the systematic application of these automorphism frame or hybrid automorphism/1-form frames in the search for new rank 2 CB geometries.
The remainder of the paper will be devoted to developing a non-trivial example demonstrating the usefulness of the automorphism frame in the case of a family of curves with enhanced automorphism group.

Also, we have included, for the interested reader, a discussion of the construction of rank 1 scale-invariant CBs from the point of view of automorphism frames in appendix \ref{appg1}.

\subsection{Application to isotrivial CB geometries}

A particularly simple class of CB geometries are the \emph{isotrivial} ones \cite{Cecotti:2021ouq}.
These are geometries where the low energy EM coupling $\t$ is constant over the CB.
The family of RSs encoding such a geometry is necessarily locally constant over $\CB^*$, and so the only EM duality monodromies that the family can have are those in the image of the automorphism group in $\Sp(4,\Z)$, i.e., in the finite subgroup of $\Sp(4,\Z)$ which fixes $\t$.
Thus these are a natural class of geometries which are expected to have enhanced automorphism groups.

All $\cN{=}3$ SCFTs are isotrivial \cite{Aharony:2015oyb,Cordova:2016xhm,Argyres:2019yyb}.
The CBs of $\cN{=}4$ sYM theories have long been known \cite{Seiberg:1997ax} to be orbifolds by the Weyl groups of their gauge algebras, and a class of $\cN{=}3$ theories have CBs which are orbifolds by complex crystallographic reflection groups \cite{Caorsi:2018zsq, Aharony:2016kai, Argyres:2019yyb, Kaidi:2022lyo}.
There are probably also both orbifold and non-orbifold isotrivial CB geometries at rank 2 with just $\cN{=}2$ supersymmetry just as there are at rank 1.
Thus there is a large class of isotrivial geometries that may be suitable targets for an automorphism frame approach to constructing their SW curves.%
\footnote{It was shown in \cite{Cecotti:2021ouq} that another large class of theories --- those with ``characteristic dimension" not equal to 1 or 2 --- are not only isotrivial but also have diagonal $\t$.  These theories are described by ``split" RSs, so are not considered here.}

\paragraph{$\cN{=}4$ CBs.}

Perhaps the best-understood of these CB geometries are those of $\cN{=}4$ sYM theories.
The CB of $\cN{=}4$ sYM with simple gauge algebra $\fg$ (and with no 2-form global symmetry \cite{Bourget:2018ond, Argyres:2018wxu}) is given by the orbifold $\C^r/\text{Weyl}(\fg)$ where the Weyl group acts by complex reflections on $\C^r$.  
(We are just looking at an $\cN=2$ CB ``slice'' of the full $3r$-dimensional CB of the $\cN{=}4$ theory.)  
In particular, if $\r\in\text{Weyl}(\fg)$ is a real reflection group element, which acts on $\R^r$ as an element of $\GL(r,\Z)$, then $\r$ acts on $\C^r = \R^r \oplus \R^r$ as
\begin{align}\label{WeylZact}
M_\r := \bpm \r & 0 \\ 0 & \r^{-t} \epm \in \Sp(2r,\Z)
\end{align}
since it preserves the symplectic form $J = \bspm 0 & -1\\1&0\espm$.
Explicitly, $\text{Weyl}(\fg)$ is generated by reflections through hyperplanes orthogonal to simple roots, $\a_i$, of $\fg$, 
\begin{align}\label{WeylRact}
\r_i : \ v &\mapsto v - 2 \frac{(\a_i,v)}{(\a_i,\a_i)} \a_i ,
\end{align}
where $(\cdot,\cdot)$ is the Killing form on the dual Cartan subalgebra.  So, acting on a basis of simple roots, 
\begin{align}\label{WeylZact2}
(\r_i)^j_k = \d^j_k - 2 \frac{(\a_k,\a_i)}{(\a_i,\a_i)}\d^j_i \in \GL(r,\Z).
\end{align}
The associated $M_i \in \Sp(2r,\Z)$ are given in \eqref{WeylZact}.

The special coordinates are flat complex coordinates on the CB and the low energy coupling matrix, $\t_{ij}$, is constant on the CB, and is fixed by all the $M_i$.  If we write $M=\bspm A&B\\C&D\espm$ in $r\times r$ block form, then $M\circ \t = (A\t+B)(C\t+D)^{-1}$, so $\t=M_i\circ\t$ becomes
\begin{align}\label{}
\t = \r_i \t \r_i^t \quad \forall i .
\end{align}
Using the definition of $\r_i$ it then follows that for simple $\fg$
\begin{align}\label{N4cplg}
\t_{ij} = z (\a_i,\a_j)
\end{align}
for arbitrary $z\in \C$.   The proportionality factor, $z$, is a function of the exactly marginal coupling of the $\cN{=}4$ theory.  

The simple rank 2 cases have $\fg= A_2$, $BC_2$, and $G_2$.  
Using \eqref{N4cplg} we find
\begin{align}\label{rk2N4cplg}
\t_{A_2} &= z
\bpm \ph{-}2 & -1 \\ -1 & \ph{-}2 \epm, &
\t_{BC_2} &= z
\bpm \ph{-}2 & -2 \\ -2 & \ph{-}4 \epm, &
\t_{G_2} &= z
\bpm \ph{-}2 & -3 \\ -3 & \ph{-}6 \epm. 
\end{align}
By conjugating $\t_{A_2}$ with $G_R$, $\t_{BC_2}$ with $G_T\circ G_S$, and $\t_{G_2}$ with $G_R\circ G_T\circ G_S$, where $G_{R,S,T}\in\Sp(4,\Z)$ are defined in appendix \ref{appZ} and their actions on $\t$ are given in \eqref{GRSTact}, we see that these CB couplings are EM duality equivalent to
\begin{align}\label{rk2N4cplg2}
\t_{A_2} &\simeq \t_{G_2} \simeq  z \bpm 2 & 1 \\ 1 & 2 \epm, &
\t_{BC_2} &\simeq z \bpm 2 & 0 \\ 0 & 2 \epm.
\end{align}
Comparing to the $\t_{ij}$ column of table \ref{tabA}, this shows that the $\fg=A_2$ and $\fg=G_2$ curves have automorphism group $D_6$, while the $\fg=BC_2$ curve must be split.

The above discussion of rank 2 $\cN{=}4$ CB geometries is incomplete.
The reason is that there can be representations of the Weyl group action on the CB which are inequivalent over the integers but are equivalent over the rationals.
Integrally inequivalent representations correspond to distinct (inequivalent) SK structures of the CB.
It turns out that at rank 2, Weyl$(A_2)$ and Weyl$(G_2)$ each have just one integral representation, given in \eqref{WeylZact} and \eqref{WeylZact2}, whereas Weyl$(BC_2)$ has two.
In addition to the split one we presented, there is another, non-split, SK structure; see \cite{ABGLW22} for a general discussion.
This non-split structure is implicitly given in \cite{gottschling1961fixpunkte, gottschling1961fixpunktuntergruppen, Argyres:2019ngz}, and implies that the fixed coupling for $BC_2$ is $\t'_{BC_2}= \bspm 2z & z \\ z & \frac12(z-z^{-1})\espm$.
From the EM duality equivalences derived in appendix \ref{appZ} and summarized in table \ref{tabGNU}, we see that $\t'_{BC_2} \simeq \frac12 \bspm z&1\\1&z\espm$, so its associated curve has automorphism group $D_4$.

\section{An example: the $\su(3)$ $\cN{=}4$ sYM curve}
\label{sec4}

We will now show how to use an appropriate automorphism frame to solve the SK integrability conditions for the curve and 1-form basis describing the $\fg=\su(3)$ $\cN{=}4$ sYM CB.
We showed at the end of the last section that such a curve must have automorphism group $D_6$.
Furthermore, there is just a 1-parameter family of such RSs, and the $\cN{=}4$ sYM theory comes in a 1-parameter family labeled by its exactly marginal coupling, $z$.
Thus, for each value of the coupling the SW curve has a fixed complex modulus, so can be described by a single fixed curve with $D_6$ automorphism.

The family, $\S(c)$, of genus-2 RSs with automorphism group $D_6$ can be parameterized as the hyperelliptic plane curves
\begin{align}\label{D6crv}
    y^2 = x^6+c x^3+1,
\end{align}
with $c \in \C$.%
\footnote{The curve degenerates when $c\in \{\pm2,\infty\}$. 
As $c\to\pm2$ three pairs of branch points collide, and correspond physically to weak coupling limits.
As $c\to\infty$ two triplets of branch points collide which gives a split RS with modulus $\t \simeq e^{i\pi/3} I_2$, which does not correspond to any weak coupling limit.} 
For a flat CB geometry, we take $c$ to be constant on the CB, so it is interpreted as (some function of) the dimensionless coupling constant of the corresponding SCFT.

The automorphism group of $\S(c)$ is the subgroup of reparametrizations $G_\aut \subset \GL(2,\C)/\Z_3$ which permute the points of the curve. 
Direct examination of the curve shows that it is the group generated by the three elements
\begin{align}\label{D6GLgen}
    g_1 &\doteq \bpm -1 & 0 \\ 0 & -1 \epm, &
    g_2 &\doteq \bpm 1 & 0 \\ 0 & \z^2 \epm, & 
    g_3 &\doteq \bpm 0 & 1 \\ 1 & 0 \epm , 
\end{align}
where 
\begin{align}
    \z \doteq e^{2\pi i/3} .
\end{align}
These generate the dihedral group of order 12, $D_6$.
As an abstract group, $D_6\simeq S_3\times \Z_2$.
It is related to the Weyl groups $D_6 \simeq \text{Weyl}(G_2) \simeq \text{Weyl}(A_2) \times \Z_2$.

\subsection{$\Sp(4,\Z)$ representation of the $D_6$ automorphism group}

\begin{figure}[ht]
\centering
\begin{tikzpicture}[auto]
\begin{scope}[scale=.7,xshift=0cm,very thick,decoration={
    markings, mark=at position 0.5 with {\arrow{latex}}}] 
	\node [circle,scale=2,draw,fill,inner sep=1pt] 
	at (2,0) [] (3) {};
	\node [circle,scale=2,draw,fill,inner sep=1pt] 
	at (5,0) [] (5) {};
	\node [circle,scale=2,draw,fill,inner sep=1pt] 
	at (-1,1.73) [] (3z) {};
	\node [circle,scale=2,draw,fill,inner sep=1pt] 
	at (-2.5,4.33) [] (5z) {};
	\node [circle,scale=2,draw,fill,inner sep=1pt] 
	at (-1,-1.73) [] (3z2) {};
	\node [circle,scale=2,draw,fill,inner sep=1pt] 
	at (-2.5,-4.33) [] (5z2) {};
	\draw[postaction={decorate},red] (3) to node {$\a_0$} (5);
	\draw[postaction={decorate},blue] (3) to node [swap] {$\g_2$}  (5z);
	\draw[postaction={decorate},red] (3z) to node {$\a_1$}  (5z);
	\draw[postaction={decorate},blue] (3z) to node [swap] {$\g_0$}  (5z2);
	\draw[postaction={decorate},red] (3z2) to node {$\a_2$}  (5z2);
	\draw[postaction={decorate},blue] (3z2) to node [swap] {$\g_1$}  (5);
\end{scope}
\end{tikzpicture}
\caption{A choice of paths between branch points on the complex $x$-plane.}
\label{figcyc}
\end{figure}
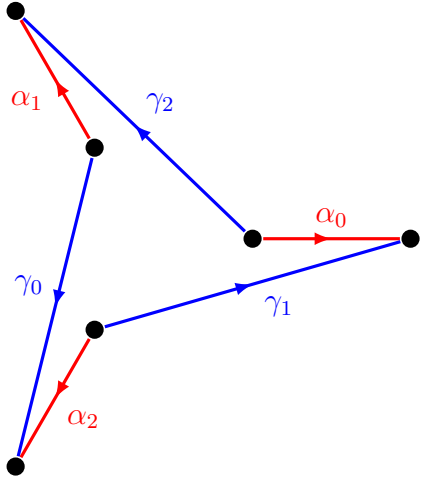

First pick a canonical basis of cycles on the curve $\S(c)$, which can be thought of as a 2-sheeted cover of the complex $x$-plane branched over the 6 zeros of the right hand side of \eqref{D6crv}.
A symmetrical set of cycles are given in figure \ref{figcyc}, where the black dots are the 6 branch points in the $x$-plane.
In drawing this figure we have assumed $c\in \R^+$ so the $\a_j$ paths are along the $\z^j$ phase lines.
The marked paths show only half of each associated cycle: the other half either returns on the other sheet or loops back on the other side of the cut (depending on where you place the cuts).
It is immediate that 
\begin{align}
    0 &= \sum_j \a_j = \sum_j \g_j , &
    &\text{and}&
    0 &= \a_j \cdot \a_k = \g_j \cdot \g_k &
    &\text{for all $j,k$,}
\end{align}
where the dot means intersection. 
Picking uniform conventions for how the $\a_j$'s and $\g_j$'s loop around the branch points (not shown in the figure) it is not too hard to see that also
\begin{align}
    \a_j \cdot \g_k = \d_{j,k-1} - \d_{j,k+1} ,
\end{align}
where the indices are understood mod 3. 
From this it follows that a canonical basis of cycles is $\{ \b_1 , \b_2 , \a_1 , \a_2 \}$ with $\b_1 \doteq \g_2$ and $\b_2 \doteq -\g_1$.

The action of the three generators on this canonical basis determines the $\Sp(4,\Z)$ representation, $N$, of the automorphism group.  
The first generator, $g_1$, interchanges the two sheets without transforming $x$, so it acts by taking every cycle to its negative. 
The second generator leaves the sheets unchanged, but rotates $x\mapsto \z x$, so takes $\a_k\mapsto \a_{k+1}$ and $\g_k\mapsto \g_{k+1}$.   Finally, the third generator, $g_3$, interchanges each ``outer" branch point with its corresponding ``inner" branch point while at the same time taking the $\z^j$ cut to the $\z^{-j}$ cut.  
This leaves the $\a_0$ and $\g_0$ cycles unchanged, but interchanges $\a_1\leftrightarrow \a_2$, and $\g_1\leftrightarrow -\g _2$. 
Thus the $\Sp(4,\Z)$ representation is (where the large ``0" and ``1" entries denote $2\times2$ blocks)
\begin{align}\label{D6Spgen}
    N(g_1) &= -\bpm 1 & 0 \\ 0 & 1 \epm ,&
    N(g_2) &= \bpm \bsm-1&1\\-1&0\esm & 0 \\
                0 & \bsm0&1\\-1&-1\esm \epm,&
    N(g_3) &= \bpm \bsm0&1\\1&0\esm & 0 \\
                0 & \bsm0&1\\1&0\esm \epm,
\end{align}
where the rows and columns are in the order $\{\b_1, \b_2, \a_1, \a_2 \}$.
$N(g_2)$ and $N(g_3)$ together fix $\t \propto \bspm 2&1\\1&2\espm$, , which is the $\cN=4$ $\su(3)$ and $G_2$ sYM coupling matrix.
$g_2$ and $g_3$ generate the Weyl$(A_2) \simeq S_3 \subset D_6$ subgroup.

\subsection{SK integrability conditions in the automorphism frame}
\label{SKint sec}

Now we solve the SK integrability conditions which are reviewed in some detail in appendix \ref{appA}. Specifically, consider \eqref{SKint1} in the automorphism frame where the curve coefficients are fixed to \eqref{D6crv}, but the $a_i^j$ coefficients of the holomorphic 1-form basis \eqref{1fbasis2} are undetermined.

Since the scaling dimensions of the $u_i \doteq (u_1,u_2) \doteq (u,v)$ CB coordinates for the $\su(3)$ $\cN{=}4$ sYM theory are $\D_u=2$ and $\D_v=3$, we have from \eqref{CBdims} that $p=2$, $q=3$, and $R=1$.
Going to the dimensionless coordinates defined in \eqref{wdef} and \eqref{scaling1} gives
\begin{align}\label{su3scaling}
    w & \doteq u^{1/2} v^{-1/3}, &
    \a^j_k(w) & \doteq v^{j/3} a^j_k, &
    \b_s(w) & \doteq v^{4/3} b_s, &
    \g_n &\doteq \d_{n,6} + c \,\d_{n,3} + \d_{n,0},
\end{align}
where we have used that $\D_x=\D_y=0$ since the curve coefficients are dimensionless numbers in this frame.
We should think of the dimensionless CB coordinate, $w$, as taking values in a projective plane which is a degree-6 function over the CB ramified at two subvarieties.
In particular, the 6 points satisfying $w^6 = t \in \C^*$ correspond to the $u^3=t v^2$ subvariety of the CB, the $w=0$ point corresponds to the $u=0$ line in the CB, and the $w=\infty$ point to the $v=0$ line. 

The eight integrability equations \eqref{SKint1} simplify considerably and  immediately imply, since they have to hold for all values of the constant $c$, that the only solution for the $\b_s$'s are that they all vanish.
Then the integrability equations reduce to just 2 equations of the same form, 
\begin{align}\label{ICsu3}
    0= w (2w \a^u_j)' + (3\a^v_j)' , \qquad j=1,2,
\end{align}
where the prime denotes derivative with respect to $w$.

These equations are not very restrictive by themselves; e.g., they essentially just determine the $\a^v_j$ in terms of arbitrary $\a^u_j$.
Clearly additional conditions are needed to specify solutions.
These conditions are precisely the 1-form basis monodromies coming from the automorphism twist.
Specifying these boundary conditions requires knowledge of the singularity locus of the CB, as well as a specific choice of automorphism twist $\a: \pi_1(\CB^*) \to \aut$.
We will show how these boundary conditions are determined for the $\su(3)$ $\cN{=}4$ sYM theory in the next subsection.

It will be convenient first to reformulate the integrability conditions in a way that makes boundary conditions they satisfy under analytic continuation more apparent.
The integrability equations are essentially necessary conditions for the periods of the $\Om_j$ 1-form basis to be derivatives of the special coordinates $s_j$, defined in \eqref{periodmat}.
Since $\b_s=0$, the integrability condition implies the 1-form basis (and not just their periods) are the derivatives of the SW 1-form $\Lambda$ given in \eqref{Zd}, $\Om_j = \del_j \Lambda$.
By \eqref{Zd} and going to scale-invariant variables, 
\begin{align}\label{lambjdef}
    \Lambda &\doteq v^{1/3} \l_j \w_j, &
    \l_j &= 2w^2\a^u_j + 3 \a^v_j ,
\end{align}
where we have decomposed $\Lambda$ in the canonical 1-form basis.
\eqref{lambjdef} can be inverted as
\begin{align}\label{defsj}
2w\a^u_j &= f(\l_j,w), &
3\a^v_j &= \l_j - w f(\l_j,w) ,
\end{align}
for some function $f$ implicitly defined by the first equation.
The integrability equations \eqref{ICsu3} then imply that $f$ and the $\l_j$ are related by 
\begin{align}\label{ICsu3 2}
    \l'(w) = f(\l(w), w) . 
\end{align}
This can, in turn, be solved by assuming that $\l(w)$ is defined implicitly by an equation
\begin{align}\label{ICsu3 3}
    \Theta(\l,w)=0
\end{align}
for some $\Theta$.  
Differentiating \eqref{ICsu3 3} with respect to $w$ then gives \eqref{ICsu3 2} with
\begin{align}\label{ICsu3 4}
    f(\l,w) \doteq -\frac{\del\Theta / \del w}{\del\Theta / \del\l} .
\end{align}

This reformulation of the SK integrability equations \eqref{ICsu3} as the single functional relation \eqref{ICsu3 3} will prove useful below.
Notice that we have dropped the index on $\l_j$ in \eqref{ICsu3 2}--\eqref{ICsu3 4}.
The idea behind this is that \eqref{ICsu3 3} is a polynomial relation between $\l$ and $w$ whose solution, $\l(w)$, is branched over the $w$-sphere.
Then the $j$ index on $\l_j$ will refer to different branches of this function, and the periodicity conditions on the 1-form basis matrix on the cut CB will be associated to the deck transformation of the branched covering $\l$ of the $w$-sphere.
Since these periodicity conditions reflect the automorphism twist, we should expect that there will be a morphism $\aut(\l) \to \im \,\a$ from automorphism group of the $\l$ covering to the image of the automorphism twist.
We will see an explicit example of this in section \ref{ssorbi}.

\subsection{Boundary conditions on 1-form coefficient functions} 

The CB singular locus for the $\cN{=}4$ $\su(3)$ theory has only a single component which we can put at $u^2=v^3$ by appropriately rescaling the $u$ or $v$ coordinate.
This follows either from a physical analysis of the CB Higgs mechanism at weak coupling, or from the orbifold description of the CB.
The EM duality monodromy around this singularity squares to the identity.
This follows from the fact that at the singularity the massless $\cN{=}4$ $\U(1)^2$ effective gauge theory at regular points is enhanced to an $\cN{=}4$ $\U(1)\times\SU(2)$ theory, and the EM duality monodromy of the $\SU(2)$ factor squares to 1.
Equivalently, it follows from the fact the that $u^2=v^3$ singularity is the fixed point locus of the elements of the Weyl group which square to 1.

We now translate these conditions into monodromy conditions on $\a_i^j(w)$ on the $w$-sphere.  
Since $w \doteq u^{1/2} v^{-1/3}$, the $u^3=v^2$ singularity corresponds to the 6 points $w^6=1$.  
Also, since $w$ is branched over the $u=0$ and $v=0$ loci, the $\a_i^j$ will satisfy special regularity conditions at $w=0$ and $w=\infty$.

In particular, there are no singularities at $u=0$ or $v=0$, so $a_i^j(0,v\neq 0)$ and $a_i^j(u\neq 0,0)$ are regular.
Recalling the definition \eqref{su3scaling} of the associated scale-invariant functions $\a_i^j$, this implies that $\a_i^j$ are regular at $w=0$ and $\infty$ and that
\begin{align}\label{w0infbc}
    \a^k_j(-w) & = \a^k_j(w) \qquad \ \,\text{around}\quad w=0,\nn\\
    \a^k_j(\z w) & = \z^{2k} \, \a^k_j(w) \quad \text{around}\quad w=\infty.
\end{align}
Furthermore, $\a^k_j$ vanish as $\sim w^{-k}$ at $w=\infty$.

At the $w^6=1$ singularities the monodromies have to be selected from those in the automorphism group of the curve. 
From \eqref{D6GLgen} it follows that there are just three nontrivial $\Z_2$ monodromies $h_n \in \aut \subset \GL(2,\C)/\Z_3$: $h_n\doteq \bspm 0 & \z^n \\ \z^{-n} & 0 \espm$,  $n\in \Z_3$.
By \eqref{Gamap}, upon continuing around any of the $w^6=1$ points, the 1-form $a_i^j$ coefficient matrix suffers the monodromy $a\mapsto \det(h_n) h_n a$ for one of these $h_n$.
These imply the associated monodromies 
\begin{align}\label{w1monod}
    h_n: (\a^j_1, \a^j_2) &\mapsto -(\z^{-n} \a^j_2, \z^n \a^j_1) 
    \qquad \text{around} \quad w^6=1 .
\end{align}
Since this monodromy squares to 1, the $\a^j_k$ have at most square-root branch cuts at the $w^6=1$  points.  

From \eqref{defsj} it follows that $\l_j$ is regular everywhere except at the roots of $w^6=1$ where it has square root branch points, and at $w=\infty$ where it has a simple pole $\sim w$, and has periodicities or monodromies
\begin{align}\label{lamb0infbc}
    \l_j(-w) & = \l_j(w) &
    &\text{around}& w&=0,\nn\\
    \l_j(\z w) & = \z \, \l_j(w) &
    &\text{around}& w&=\infty,\nn\\
    h_n: (\l_1,\l_2) &\mapsto -(\z^{-n} \l_2, \z^n \l_1) &
    &\text{around}& w^6 &= 1 .
\end{align}
It is not clear to us whether these conditions (perhaps made more precise by specifying which $h_n$ monodromy is associated to each root of $w^6=1$) are enough to determine $\l_j$ up to equivalences.
But we can use our knowledge of the orbifold description of the $\su(3)$ $\cN{=}4$ CB to determine $\l_j$ satisfying these conditions in a simple way, as we now show.

\subsection{Solution for the 1-form basis from the orbifold description}
\label{ssorbi}

At least in the case of orbifold CBs, where the special coordinates are the flat coordinates, the holomorphic action of the finite orbifold group implies that the flat coordinates are algebraically related to the invariant CB coordinates.

The $\su(3)$ $\cN{=}4$ CB is given by $\C^2/\text{Weyl}(A_2)$ where Weyl$(A_2)$ acts by complexified reflections \eqref{WeylRact}. 
The orbifold invariant $(u,v)$ CB coordinates are given by the holomorphic Weyl-invariant combinations of the flat coordinates.
The flat coordinates in isotrivial geometries are linearly related to the special (or dual special) coordinates, \eqref{periodmat} and \eqref{specialcoords} \cite{Argyres:2019yyb}.
These coordinates are periods of the SW 1-form \eqref{Zd}, $\Lambda = \Lambda_j \w_j$, so are linear combinations of the $\Lambda_j$.
The Weyl action is equivalent to permutations of 3 linear combinations, $\Lambda_n$, $n=0,1,2$, which sum to 0, so $u$ and $v$ are the quadratic and cubic invariants of these three $\Lambda_n$'s. 
Therefore the $\Lambda_n$'s are the roots of the cubic polynomial 
\begin{align}\label{Qdef}
    Q(s,u,v) \doteq \Lambda^3-3u \Lambda+2v ,
\end{align}
whose discriminant with respect to $\Lambda$ is $108(u^3-v^2)$.
Here we've chosen the normalizations of $u$ and $v$ so that the zero of the discriminant is at $u^3=v^2$. 

In terms of scale-invariant variables $w$, $\l_n \doteq v^{-1/3} \Lambda_n$, and $\Theta \doteq v^{-1} Q$, $Q=0$ becomes
\begin{align}\label{Theta0}
    0 = \l^3 - 3 w^2 \l + 2 = \Theta(\l,w) ,
\end{align}
with disc$_\l\Theta = 108(w^6-1)$.
Note that \eqref{Theta0} implies the identities
\begin{align}\label{lambidents}
    0 &= \l_0+\l_1+\l_2, &
    -3w^2 &= \l_0\l_1 +\l_1\l_2 +\l_2\l_3, &
    -2 &= \l_1\l_2\l_3 .
\end{align}
The periodicities and monodromies of $\l(w)$ can be read off from \eqref{Theta0}.
The roots are even functions of $w$ for $|w|<1$, and are finite everywhere except at $w=\infty$ where two roots diverge as $\sim \pm \sqrt3 \, w$ and one root vanishes as $\sim (2/3) w^{-1}$. 
Also, since the discriminant vanishes to first order at $w^6=1$, only pairs of roots collide at those points, giving two sheets with a square root branch point, and a regular third sheet at each root of $w^6=1$.  
In particular, upon traversing a small counterclockwise loop around the $w^6=1$ branch points, the sheets undergo the monodromies
\begin{align}\label{lambmonod}
    \l_{n-1} &\mapsto \l_{n+1}, &
    \l_n &\mapsto \l_n, &
    \l_{n+1} &\mapsto \l_{n-1}, &
    &\text{around}& w^2 &= \z^{-n}, 
\end{align}
where the $n$ index is understood mod 3.

Then, from \eqref{Theta0}, \eqref{ICsu3 4} and \eqref{defsj} we find three solutions to the integrability equation,
\begin{align}\label{alpha solution}
    \til\a^u_n &\doteq \frac{\l_n}{\l_n^2-w^2}, &
    \til\a^v_n &\doteq \frac{-(2/3)}{\l_n^2-w^2} .
\end{align}
Note that although the three $\l_n$'s are not linearly independent by \eqref{lambidents}, the $\til\a_n$'s are.
The tildes indicate that this is not quite the final solution for the 1-form basis coefficients, since we have not yet imposed the monodromy condition \eqref{w1monod} around the $w^6=1$ points.
So we look for 
\begin{align}\label{atil to a}
    \a^k_j = C^n_j \til\a^k_n
\end{align}
for some constant $2\times3$ matrix $C$ such that $\a^k_j$ have the desired monodromy.
The monodromy of the $\til\a^k_n$ are simply those  of the $\l_n$, \eqref{lambmonod}, while the desired monodromies are the $h_m$ monodromies \eqref{w1monod} in the automorphism group for some assignment of $m\in\Z_3$ to the $w^6=1$ branch points.
It is a short exercise in algebra to see that there are two possible assignments of $h_m$'s which give solutions for $C$ (unique up to overall normalization) satisfying these requirements,
\begin{align}\label{C soln}
    C = \bpm -1 & -\z^{2n} & -\z^n \\ \z^n & \z^{2n} & 1 \epm
    \ \ \text{for}\ n = \pm 1, 
    \qquad \text{if $w^2=\z^m$ has monodromy $h_{n(m+1)}$.}
\end{align}
Here the exponents and subscripts are mod 3.  This is also a solution of the monodromy constraint for $n=0$, but in this case the rows of $C$ are not linearly independent, so the resulting 1-form basis is degenerate.
We can take $n=1$ without loss of generality, since the $n=-1$ solution is equivalent as it consists of picking for $\z$ the other primitive cube root of unity.

Plugging this into \eqref{atil to a} and using the identities \eqref{lambidents} judiciously, one finds
\begin{align}\label{1fbasis solution}
\a^u_1 &= 
\frac{+(\l_0 {-} \l_1) (\l_0 \l_1 {+} w^2) (\l_2^2 {-} w^2)
-\z (\l_1 {-} \l_2) (\l_1 \l_2 {+} w^2) (\l_0^2 {-} w^2)}
{4 (1-w^6)}
, \nn\\
\a^u_2 &= 
\frac{-\z(\l_0 {-} \l_1) (\l_0 \l_1 {+} w^2) (\l_2^2 {-} w^2)
+ (\l_1 {-} \l_2) (\l_1 \l_2 {+} w^2) (\l_0^2 {-} w^2)}
{4 (1-w^6)}
, \nn\\
\a^v_1 &=
\frac{-(\l_0^2 {-} \l_1^2) (\l_2^2 {-} w^2)
+\z (\l_1^2 {-} \l_2^2) (\l_0^2 {-} w^2)}
{6 (1-w^6)}
, \nn\\
\a^v_2 &= 
\frac{+\z (\l_0^2 {-} \l_1^2) (\l_2^2 {-} w^2)
- (\l_1^2 {-} \l_2^2) (\l_0^2 {-} w^2)}
{6 (1-w^6)} . 
\end{align}
This is the solution --- unique up to overall normalization and overall permutation of the $\l_n$ --- for the $\cN{=}4$ $\su(3)$ 1-form basis coefficients in the automorphism frame.
Since the $\l_n$ are the roots of the cubic equation \eqref{Theta0}, they have explicit algebraic expressions;  however, those do not add anything to the expression given in \eqref{1fbasis solution}.

\subsection{The curve in canonical $\GL(2,\C)/\Z_3$ frame}

It is a straight forward computation to transform the automorphism frame solution \eqref{D6crv} and \eqref{1fbasis solution} to the canonical frame where the 1-form basis is the canonical one \eqref{cfa} but the curve coefficients vary meromorphically over the CB.
As explained below \eqref{cfa}, this is done by acting on the automorphism frame solution with the $\GL(2,\C)/\Z_3$ reparameterization given by $G= a^{-1} (\det a)^{1/3}$ where $a$ is the 1-form basis coefficient matrix in the automorphism frame.
Using \eqref{su3scaling}, the coefficient matrix is 
\begin{align}
a = \bpm  
v^{-1/3} \a^u_0 & v^{-1/3} \a^u_1 \\
 v^{-2/3} \a^v_0 & v^{-2/3} \a^v_1 
\epm ,
\end{align}
and then a lengthy calculation using the identities \eqref{lambidents} results in the complicated curve
\begin{align}\label{su3 crv sol}
y^2 &=  \frac1{u^3-v^2} \Bigl(
\tfrac19 \left[(c{+}2) u^3 - 4 v^2\right] x^6 
+ (c{-}2) u^2 v \,x^5
+\tfrac34 \left[(c{-}10) u^3 + 4c v^2\right] u \,x^4
\nn\\
&\qquad\qquad\quad
+\tfrac32 \left[(3c{-}10) u^3 + 2c v^2\right] v \,x^3
+\tfrac{27}{16} \left[(c{+}10) u^3 + 4 (c{-}5) v^2\right] u^2 x^2
\nn\\
&\qquad\qquad\quad
+\tfrac{81}{16} \left[(c{+}6) u^3 - 8 v^2\right] u v \,x
+\tfrac{81}{64} \left[(c{-}2) u^6 + 16 u^3 v^2 - 16 v^4\right]
\Bigr) .
\end{align}
Recall that $c$ parameterizes the exactly marginal coupling of the $\su(3)$ sYM theory.
This is the solution reported in \cite{Argyres:2022kon}.
One can check that the integrability condition \eqref{SKint1} is satisfied, though in a quite intricate way.

Its discriminant is proportional to $(c^2-4)^3 (u^3-v^2)^5$.
This is computed most easily using \eqref{Gdmap}.
It is a monomial in $u^3-v^2$, as expected since that is the only singular locus on the CB.
However, the power of $u^3-v^2$ is 5, and not 6 as is expected for the ``quantum discriminant" introduced in the CB stratification analysis of \cite{Martone:2020nsy}.


\section{Future directions}
\label{sec5}

The $\su(3)$ sYM calculation suggests a systematic approach to constructing scale-invariant SW curves with extra automorphisms:
\begin{enumerate}
    \item For a given automorphism group, aut, put the curve in the corresponding canonical form, as in table \ref{tabA} or as outlined in the automorphism frames discussion in section \ref{aut frames}.
    \item Choose a set of CB scaling dimensions, $(\D_u,\D_v)$, from the set of allowed pairs found in \cite{Caorsi:2018zsq, Cecotti22}.
    \item Choose the CB singular locus, $\cV$, as a union of the varieties in \eqref{knots}.
    \item Choose an automorphism twist $\a$, \eqref{aut twist}.
    \item Solve the SK integrability equations \eqref{SKint1} for the curve and the 1-form basis in the automorphism frame using the automorphism twist \eqref{aut twist} and \eqref{aBC} as boundary conditions.
\end{enumerate}
If no solution exists, then there is no SK geometry with the data chosen in steps 1--4.
Even if one or more solutions exist, they still may not correspond to the CBs of SCFTs:  an additional set of consistency conditions under RG flows --- analogous to those analyzed in the rank-1 case in \cite{Argyres:2015ffa, Argyres:2015gha, Argyres:2016xua, Argyres:2016xmc, Argyres:2016yzz, Argyres:2017tmj} and partially explored in higher rank in \cite{Martone:2020nsy, Argyres:2020wmq} --- need to be satisfied.
Furthermore, a given scale-invariant solution can correspond to many distinct SCFTs, as was found to be the case at rank 1 and as discussed in many examples in rank 2 in \cite{Argyres:2022lah, AM22II}.

We now discuss each of these steps in a bit more detail to give a sense of the challenges involved in turning this approach into a finite and systematic procedure.

Step 1 is clearly finite, consisting of choosing one of the 6 entries in the curve column of table \ref{tabA} with enlarged automorphism group.

Step 2 is also finite, as the allowed rank-2 CB scaling dimensions is a known set of 79 allowed pairs \cite{Caorsi:2018zsq, Cecotti22, Argyres:2022lah}.

Step 3 is not known to be finite, since $\cV$ could, in principle, have an arbitrary number of $u^q=tv^p$ sub-varieties with distinct values of $t$.
But there are strong constraints on the number and type of these singularity sub-loci which come from the CB scaling dimension and conformal central charge calculus of \cite{Martone:2020nsy}.

Step 4 is finite since $\pi_1(\CB^*)$ has been computed for every possible choice of $\cV$  \cite{Argyres:2019kpy}, and aut(RS) are relatively small finite groups.

Step 5 is a system of non-linear ODEs for the coefficients of the curve and the 1-form basis matrix, and so might be difficult to analyze.
However, the form of the canonical curves with extra automorphisms is very constrained, so most (if not all) of the curve coefficients are fixed, which turns the SK integrability conditions into a system of \emph{linear} ODEs for the 1-form basis matrix.
Furthermore, as we saw in the example in section \ref{SKint sec}, integrability conditions simplify dramatically for curves with extra automorphisms.

This computational strategy is weakest in the case of small automorphism groups $\aut=\Z_2$ (the generic case) and $\aut=D_2$ (enjoyed by a two-dimensional subspace of genus-2 curves).
But it may reasonably provide a systematic scan of all CB geometries described by curves with larger automorphism groups, and is clearly a powerful way of constructing isotrivial CB geometries.
In this context, an interesting outstanding question is whether there are non-orbifold isotrivial CB geometries at rank 2 as there are at rank 1.

The main unresolved question in this approach is whether and how the integrability equations (step 5) constrain the possible choices of singular locus (step 3).
From the relatively small number of scale-invariant rank 2 CB solutions that have been found to date, we expect that existence of solutions to the integrability equations will very tightly constrain the possible number and arrangement of the components of the singular locus.

Another interesting application of this approach may be to search for isotrivial CB geometries whose coordinate ring is not freely generated, i.e., which have complex singularities.
Such geometries are known to exist, and, indeed all rank 2 orbifold examples (with principal Dirac pairing) were constructed in \cite{Argyres:2019ngz}.
In particular, in these cases the special coordinates are given by a homogeneous non-principal ideal $I=\left\langle Q_s(\Lambda,u_j)\right\rangle$ instead of the principal ideal generated by \eqref{Qdef}.
Here $u_j$ are an over-complete set of scaling coordinates on the CB which therefore satisfy some relations, which are part of the ideal.
It would be interesting to work out how the integrability condition generalizes and what replaces the $\Theta(\l,w)=0$ and $f=- \del_w\Theta/\del_\l\Theta$ relations, \eqref{ICsu3 3} and \eqref{ICsu3 4}, in this case.

\acknowledgments It is a pleasure to thank A. Bourget, J. Distler, J. Grimminger, M. Lotito, and M. Weaver for many helpful discussions and comments. 
PCA is supported by DOE grant DE-SC0011784, and MM is supported by STFC grant ST/T000759/1.

\appendix

\section{The SK integrability condition}\label{appA}

In this section we briefly review the SK integrability conditions which played a central role in \cite{Argyres:2005pp, Argyres:2005wx,Argyres:2022lah}.
Recall that  a general basis of holomorphic 1-forms is then
\begin{align}
\Om_i := \w_j \, a_i^j ,
\end{align}
where $a_i^j$ vary locally holomorphically over the CB and the canonical basis of holomorphic 1-forms is defined to be:
\begin{align}
\w_u &:= x \, \frac{dx}{y} ,&  \w_v &:= \frac{dx}{y} .
\end{align}
The curve and basis of 1-forms determine an SK structure if they satisfy the \emph{SK integrability condition}
\begin{align}
\del_v\Om_u - \del_u \Om_v = d g,
\end{align}
where $\del_i := \del/\del u_i$, $g$ is a meromorphic function on the RS, and $d$ is the exterior derivative on the RS. As argued in \cite{Argyres:2005pp}, the most general form of the meromorphic function $g$ on $\S$ is
\begin{align}\label{}
g = \sum_{s=0}^3 b_s \frac{x^s}y .
\end{align}
Then the integrability condition becomes 8 coupled nonlinear PDEs in the $(u,v)$ for the 15 unknown functions:
\begin{align}\label{abc coeffs}
    a_i^j, \ i,j&\in\{1,2\}, &
    b_s, \ s&\in\{0,1,2,3\}, &
    c_n, \ n&\in\{0,1,\ldots,6\},
\end{align}
which parameterize the SW family of genus 2 RSs and a basis of holomorphic 1-forms on it.

These can be simplified using the scale symmetry.  
It is convenient to define the dimensionless complex coordinate,
\begin{align}\label{wdef}
w & \doteq  u^{1/p} v^{-1/q}  ,
\end{align}
on the CB, though it is important to keep in mind that $w$ is not single-valued on the CB.  
We can think of $(u,v)$ as homogeneous coordinates on the weighted $\P_{(p, q)}^1 \ni [u : v]$ where $[u:v] = \{(u,v) \sim (\l^p u,\l^q v), \ \l\in\C^* \}$.
Then a $(pq)$-fold cover of $\P_{(p, q)}^1$ is $\P^1 \ni [u^{1/p} : v^{1/q}]$ with $(u^{1/p},v^{1/q}) \sim (\l u^{1/p}, \l v^{1/q})$, and $w$ is the good local coordinate in the $v \neq 0$ patch of this $\P^1$ (and its inverse for the $u\neq 0$ patch).
In particular, a $u=0$ singular locus is at $w=0$, a $v=0$ one is at $w=\infty$, while a single $u^q=t v^p$, $t\in\C^*$, locus corresponds to the $pq$ points $w= t^{1/(pq)}$.

Assigning scaling dimensions $\D_{\cdots}$ to all quantities, we have
\begin{align}\label{}
\D(a_i^j) &= 1-\D_i-(3-j)\D_x +\D_y ,\nn\\
\D(b_s) &= 1-\D_u-\D_v-s \D_x +\D_y ,\\
\D(c_n) &= 2\D_y -n \D_x .\nn
\end{align}
Upon defining the dimensionless locally holomorphic functions of $w$,
\begin{align}\label{scaling1}
\a_j^i(w) &\doteq v^{(-1-\D_y +(3-j)\D_x +\D_i)/\D_v}  a_j^i , \nn \\
\b_s(w) &\doteq v^{(-1-\D_y +s \D_x +\D_u +\D_v)/\D_v} b_s , \\
\g_n(w) &\doteq v^{(-2\D_y +n \D_x )/\D_v} c_n , \nn
\end{align}
the integrability equations become  
\begin{align}\label{SKint1}
\sum_{j=0}^1 \Bigl(
& \bigl[ \D_u w\a^u_j+\D_v w^{1-p}\a^v_j \bigr] \g_{n+j}'
-2 \bigl[ \D_u w \, (\a^u_j)'+\D_v w^{1-p} \, (\a^v_j)' 
\bigr] \g_{n+j} \Bigr) \\
& \qquad =
p\sum_{j=0}^1 \bigl[ 2-2\D_u+(n-4+3j)\D_x \bigr] \a^u_j \g_{n+j} +
p\D_v \sum_{s=0}^3 (n+2-3s) \b_s \g_{n+2-s} .\nn
\end{align}
for $n\in \Z$ where $\g_n\equiv0$ if $n<0$ or $n>6$, and the prime denotes derivative with respect to $w$.  

\eqref{SKint1} gives 8 coupled ordinary differential equations for the 15 unknown functions $\a_i^j$, $\b_s$, and $\g_n$.
But, as we discuss in section \ref{secBS}, these 15 functions are subject to a 4-parameter family of coordinate reparameterizations which leave the form of the integrability condition (and the physics it encodes) unchanged.
Thus there are effectively only 11 independent functions characterizing a rank 2 scale-invariant SK geometry.
Furthermore, in this system of 8 ordinary differential equations, 4 of the unknown functions (the $\b_s$) enter linearly without derivatives.
So, after fixing the reparameterization invariance in some fashion, the SK integrability conditions reduce to a system of 4 nonlinear holomorphic ODEs in 7 unknowns.

\section{Automorphisms of genus 2 RSs}
\label{Autsec}

The possible automorphism groups, aut(RS), and the relationship between the algebraic form of the curve and aut(RS) is worked out for genus 2 in \cite{Igusa1960ArithmeticVO, Shaska2004EllipticSA}.
These results are shown in table \ref{tabA}.
They, and the rest of the statements in this subsection, hold only for \emph{regular} genus 2 RSs.
In particular, for the ``split" degenerate RSs described footnote \ref{split ftnt}, which are of physical interest, a separate analysis of the automorphism groups and their moduli spaces must be performed.

The first column of table \ref{tabA} gives the possible aut(RS).
The $D_n$ are the dihedral groups of order $|D_n|=2n$, and $\GL_2(3)$ is the group of $2\times 2$ matrices with entries which are integers mod 3 and whose determinant does not vanish (mod 3); it has order $|\GL_2(3)|=48$.
These groups have the presentations
\begin{align}
    D_n &= \langle \a,\b | \a^n = \b^2 = (\a\b)^2 = 1 \rangle, \nn\\
    \GL_2(3) &= \langle \a,\b | \a^3 = \b^2 = (\b\a)^8 = (\b\a)^4(\b\a^{-1})^4 = 1 \rangle.
\end{align}
Note that $D_2 \simeq \Z_2 \times \Z_2$.

In terms of its action on the curve \eqref{SWcrv}, aut(RS) can be thought of as the subgroup of $\GL(2,\C)/\Z_3$ acting as in \eqref{Gmap} which takes the curve to itself (but permutes the points of the curve).
All genus-2 RSs have a $\Z_2$ group 
\begin{align}\label{heZ2}
    \Z_2 &= \bigl\langle \, (x,y) \mapsto (x,-y) \, \bigr\rangle
\end{align}
of hyperelliptic automorphisms generated by $-1\in\GL(2,\C)/\Z_3$ defined in \eqref{G-1}.
This is a normal subgroup of aut(RS), so define the quotient group $\bar\aut \doteq \aut/\Z_2$, listed in column 2 of the table.

A convenient choice of canonical curves $y^2 = c^\aut(x)$ for each aut(RS) is shown in column 3.
Since 3 of the roots of $c(x)$ can be taken to arbitrary positions by fractional linear $\PSL(2,\C)$ transformations while preserving the binary sextic form of the curve, the forms of the curve given in the table are not unique.
They are chosen to make the $\PSL(2,\C)$ action of $\bar\aut$(RS) simple.
Column 4 gives a generating set for $\bar\aut$(RS) on the canonical curve.
$\z_n$ denotes a primitive $n$th root of unity.

The representation, $G_\aut$, of aut(RS) in $\GL(2,\C)/\Z_3$ is given by combining \eqref{heZ2} with the $\PSL(2,\C)$ action of the $\bar\aut$ generators.
A short calculation gives the generators of $G_\aut$ shown in table \ref{tabA2}.

\begin{table}[h!]
\centering
$\begin{array}{c|l}
\aut & \qquad\qquad G_\aut \ \text{generators} \\ 
\hline
\Z_2 & 
\bpm -1 & 0\\ 0 & -1 \epm ^{\ph2}
\\[4mm]
D_2 & 
\bpm -1 & 0\\ 0 & -1 \epm ,
\bpm 1 & 0\\ 0 & -1 \epm  
\\[4mm]
D_4 &
\bpm -1 & 0\\ 0 & -1 \epm ,
\bpm 1 & 0\\ 0 & -1 \epm ,
\bpm 0 & i\\ i & 0 \epm 
\\[4mm]
D_6 &
\bpm -1 & 0\\ 0 & -1 \epm ,
\bpm 1 & 0\\ 0 & \, \z_3^2 \epm ,
\bpm 0 & 1\\ 1 & 0 \epm 
\\[4mm]
\ \GL_2(3)\ \  & 
\bpm -1 & 0\\ 0 & -1 \epm ,
\bpm \z_8^3 & 0\\ 0 & \, \z_8 \epm ,
\bpm 0 & i\\ i & 0 \epm ,
{\displaystyle \frac{i}{\sqrt2}}\!\bpm 1 & -1\\ 1 & 1 \epm
\\[4mm]
\Z_{10} & 
\bpm -1 & 0\\ 0 & -1 \epm ,
\bpm 1 & 0\\ 0 & \, \z_5^4 \epm
\\[4mm]
\Z_3{\rtimes} D_4 &
\bpm -1 & 0\\ 0 & -1 \epm ,
\bpm 1 & 0\\ 0 & \, \z_6^5 \epm ,
\bpm 0 & i\\ i & 0 \epm
\end{array}$
\caption{\label{tabA2} 
Generating elements of $G_\aut \subset \GL(2,\C)/\Z_3$ representations of the automorphism groups of genus-2 RSs.}
\end{table}

The fifth column of table \ref{tabA} gives the complex modulus of the curve as an element $\t_{ij}$ of the rank 2 Seigel half space.
Recall that the rank 2 Seigel half space is the 3-complex-dimensional manifold of symmetric $2\times2$ complex matrices, $\t_{ij}$, with positive definite imaginary part.
Two points in the Seigel half space represent equivalent complex structures of the RS iff they are related by the usual (M\"obius-like) action of the Seigel modular group, $\Sp(4,\Z)$.
The $\t_{ij}$ given in the table is a representative of this equivalence class chosen as described in appendix \ref{appZ}. 
The moduli parameters $c_n$ which appear in the curves are some $\Sp(4,\Z)$ modular functions of the $\t_{ij}$ which we leave unspecified.  
(They are analogs of the $\SL(2,\Z)$ modular $\l$ function of elliptic curves which appear in the rank 1 case.)

Representations of $\aut$ in $\Sp(4,\Z)$ can be derived either by finding the finite subgroup of $\Sp(4,\Z)$ fixing the corresponding $\t$ in table \ref{tabA}, or by choosing a suitable canonical 1-cycle homology basis of the curve and computing how it transforms under the $\GL(2,\C)/\Z_3$ reparameterizations in $\aut$, also given in table \ref{tabA}.
The first method is carried out in \cite{gottschling1961fixpunkte, gottschling1961fixpunktuntergruppen}, which, together with the $\Sp(4,\Z)$ equivalences described in appendix \ref{appZ} below, gives the full list for all automorphism groups (including those of split RSs, not discussed here).

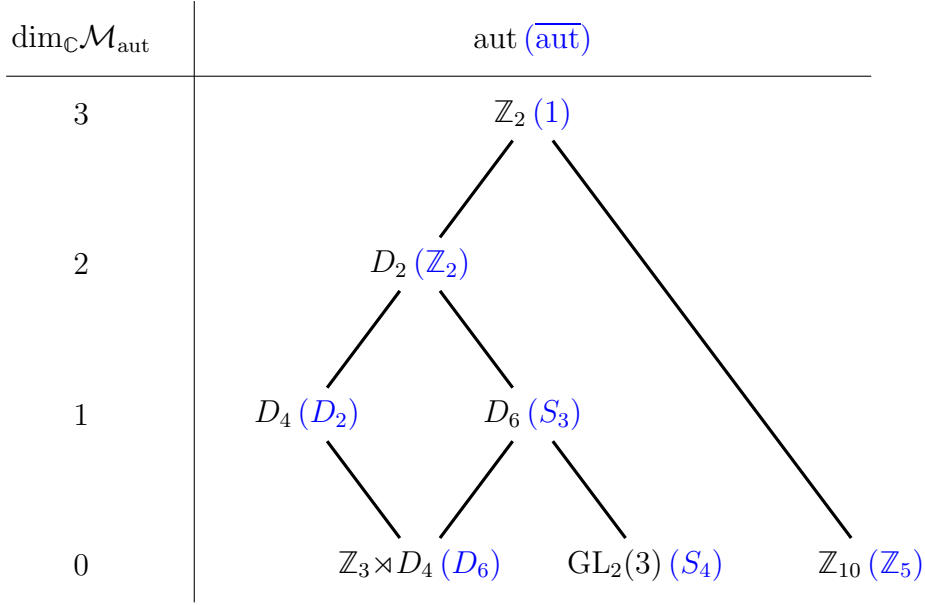
\begin{figure}[ht]
\centering
\begin{tikzpicture}
\node at (0,7) {dim$_\C \cM_\aut$};
\node at (0,6) {$3$};
\node at (0,4) {$2$};
\node at (0,2) {$1$};
\node at (0,0) {$0$};
\draw (-1,6.5) -- (10.5,6.5);
\draw (1.5,7.5) -- (1.5,-0.5);
\node at (6,7) {$\aut\, \blue{(\bar\aut)}$};
\node (Z2) at (6,6) {$\Z_2\, \blue{(1)}$};
\node (D2) at (4.5,4) {$D_2\, \blue{(\Z_2)}$};
\node (D4) at (3,2) {$D_4\, \blue{(D_2)}$};
\node (D6) at (6,2) {$D_6\, \blue{(S_3)}$};
\node (Z3D4) at (4.5,0) {$\Z_3{\rtimes}D_4\, \blue{(D_6)}$};
\node (Z10) at (10.5,0) {$\Z_{10}\, \blue{(\Z_5)}$};
\node (GL23) at (7.5,0) {$\GL_2(3)\, \blue{(S_4)}$};
\draw[very thick] (Z2) -- (D2);
\draw[very thick] (D2) -- (D6);
\draw[very thick] (D2) -- (D4);
\draw[very thick] (D4) -- (Z3D4);
\draw[very thick] (D6) -- (GL23);
\draw[very thick] (D6) -- (Z3D4);
\draw[very thick] (Z2) -- (Z10);
\end{tikzpicture}
\caption{Poset of moduli spaces of genus-2 RSs with given automorphism groups.}
\label{figA}
\end{figure}

The moduli space of regular genus 2 RSs is 3-complex-dimensional and there is a set of irreducible complex subvarieties $\cM_{\rm aut(RS)}$ of the moduli space on which the automorphism group of the RS is enhanced to aut(RS).%
\footnote{The irreducibility of the $\cM_\aut$ follows from the classification of \cite{gottschling1961fixpunkte, gottschling1961fixpunktuntergruppen} of fixed loci of the $\Sp(4,\Z)$ action on the rank 2 Seigel half-space, discussed in appendix \ref{appZ}.}
The complex dimension of these subvarieties is given in the final column in the table.
This set of subvarieties is partially ordered by inclusion, where $\cM_{\aut_1} \supset \cM_{\aut_2}$ implies $\aut_1 \subset \aut_2$.
The set of possible aut(RS), their partial ordering by inclusion, and the dimension of the subvariety, $\cM_\aut$, of the moduli space of genus 2 RSs on which they are realized are shown in figure \ref{figA}.

\section{Genus 2 moduli in an $\Sp(4,\Z)$ fundamental domain}
\label{appZ}

The complex modulus of a genus 2 RS is $\t_{ij} \in \cH_2$, where $\cH_2$ is the rank 2 Seigel half space, the 3-complex-dimensional space of $2\times2$ symmetric complex matrices with positive-definite imaginary parts.  
$\Sp(4,\Z)$ has a natural Mobius-like action on $\cH_2$, described below, and any two points of $\cH_2$ differing by an $\Sp(4,\Z)$ transformation are moduli of the same RS.
The boundaries of $\cH_2$ correspond to degenerations of the RS, as does the subvariety
\begin{align}\label{split}
    \t &= \bpm z_1 & 0 \\ 0 & z_2 \epm, &
    &\text{with}&
    z_{1,2} &\in \C ,
\end{align}
and its images under $\Sp(4,\Z)$.
We call the degeneration limit of genus 2 RSs with modulus in \eqref{split} ``split RSs".
They are genus 2 RSs which are pinched to form a bouquet of two genus 1 RSs.

An element $M\in\Sp(4,\Z)$ is an integer $4\times4$ matrix satisfying $MJM^t=J$ where $J$ is a principal integral symplectic form.
Write $M$ in terms of $2\times2$ blocks as $M = \bspm A & B\\ C&D\espm$ in a basis in which $J = \bspm 0 & 1\\-1&0\espm$.
Then $MJM^t=J$ is equivalent to $C^tA$ and $B^tD$ symmetric, and $A^t D - C^t B = 1$.
The action of $M$ on $\cH_2$ is given by
\begin{align}\label{mobius}
    M: \t \to M\circ\t := (A\t+B)(C\t+D)^{-1} .
\end{align}
Note that the $\Z_2 := \{\pm I\}$ center of $\Sp(4,\Z)$ acts trivially on $\cH_2$.

A standard fundamental domain, $\cF_2 \subset \cH_2$, of this action is defined by the inequalities \cite{klingen1990introductory} 
\begin{align}\label{F2dom}
    & |x_k| \leq \tfrac12 \qquad \text{for}\ k=1,2,3, &
    & 0 \leq 2y_3 \leq y_1 \leq y_2 , \nn\\
    & |z_k| \geq 1 \qquad \text{for}\ k=1,2, &
    & |z_1+z_2-2z_3 \pm 1| \geq 1 , \nn\\
    & |\det (Z+S)| \geq 1 ,
\end{align}
where we parameterize
\begin{align}
    \cH_2 \ni \t &:= \bpm z_1 & z_3\\z_3 & z_2 \epm, & & \text{with} &
    z_k &:= x_k + i y_k,
\end{align}
and $\pm S \in \{ \bspm0&0\\0&0\espm, \bspm1&0\\0&0\espm, \bspm0&0\\0&1\espm, \bspm1&0\\0&1\espm, \bspm1&0\\0&-1\espm, \bspm0&1\\1&0\espm, \bspm1&1\\1&0\espm, \bspm0&1\\1&1\espm \}$.
Points on the boundary of $\cF_2$ defined by \eqref{F2dom} are identified by $\Sp(4,\Z)$ transformations.

The representative $\t_{ij}$ moduli for the various subvarieties with extra automorphisms in the last column of table \ref{tabA} are chosen to lie in the $\cF_2$ fundamental domain.
But only the first and last entry in that table are generically in the interior of $\cF_2$:  all the others are in the boundary of $\cF_2$.
Since the boundaries of $\cF_2$ are identified by $\Sp(4,Z)$ transformations, there is some remaining ambiguity in the choice of $\t$ given in table \ref{tabA}.
The rest of this appendix discusses this ambiguity.
This will also be useful in the discussion of the $\t_{ij}$'s of the rank 2 $\cN=4$ sYM theories.

The $\t_{ij}$ moduli in subvarieties of moduli space with extra automorphisms can be extracted from \cite{gottschling1961fixpunkte, gottschling1961fixpunktuntergruppen, Ueno1973}.
They both give $\t_{ij}$'s in $\cF_2$, but compute them in different ways:
\cite{gottschling1961fixpunkte, gottschling1961fixpunktuntergruppen} determines them by classifying $\t_{ij}$ fixed by subgroups of $\Sp(4,\Z)$, while \cite{Ueno1973} computes them for degenerating 1-parameter families of genus 2 RSs.
In either case, by looking at the maximal subgroup fixing $\t_{ij}$ \cite{Argyres:2019ngz}, or by inspecting the form of the binary sextic curve as it approaches the degeneration, the associated automorphism group is easy to determine.%
\footnote{The great majority of the $\sim100$ cases recorded in \cite{gottschling1961fixpunkte, gottschling1961fixpunktuntergruppen, Ueno1973} pertain to split RSs as in \eqref{split}, and are not included in table \ref{tabA}.}

But these references give many different forms for $\t_{ij}$ for each automorphism group.
We now show when and how they are related by $\Sp(4,\Z)$ transformations.
It is useful to note that $\Sp(4,\Z)$ is generated by the elements (where the large ``0" and ``1" entries denote $2\times2$ blocks)
\begin{align}
    S &:= \bpm 0&-1\\1&0 \epm , &
    G_S &:= \bpm\bsm0&1\\1&0\esm &0\\0&\bsm0&1\\1&0\esm\epm, &
    1{\odot}S &:= \bspm 1&0&0&0\\ 0&0&0&-1\\ 0&0&1&0\\ 0&1&0&0 \espm , \\
    T &:= \bpm 1&\bsm p&r\\ r&q\esm \\0&1\epm, &
    G_R &:= \bpm\bsm1&0\\0&-1\esm &0\\0&\bsm1&0\\0&-1\esm\epm, &
    G_T &:= \bpm\bsm1&1\\0&1\esm &0\\0&\bsm1&-1\\0&1\esm\epm, \nn
\end{align}
where $p,q,r\in\Z$, which act on $\t$ as
\begin{align}\label{GRSTact}
    S\circ\t &= -\t^{-1} ,&
    T \circ\t &= \t+\bpm p&r\\ r&q\epm ,\nn\\
    G_S\circ\bpm z_1 & z_3\\ z_3& z_2 \epm &= 
    \bpm z_2 & z_3\\ z_3 & z_1 \epm ,&
    G_R \circ \bpm z_1 & z_3\\ z_3& z_2 \epm &= 
    \bpm z_1 & -z_3\\ -z_3& z_2 \epm ,\\
    1{\odot}S \circ \bpm z_1 & z_3\\ z_3 & z_2 \epm &= 
    \frac1{z_2} \, \bpm z_1 z_2 {-} z_3^2 & z_3\\ z_3 & -1\epm,&
    G_T \circ \bpm z_1 & z_3\\ z_3& z_2 \epm &= 
    \bpm z_1{+}z_2{+}2z_3 & z_2{+}z_3\\ z_2{+}z_3 & z_2 \epm .\nn
\end{align}
Note that the $T$ transformations imply that $\t$'s whose entries differ by integers are $\Sp(4,\Z)$ equivalent, and the $G_S$ transformation interchanges $z_1$ and $z_2$.
Using just these two equivalences easily and greatly reduces the number of different (non-split) $\t$'s to the ones shown in table \ref{tabGNU}.

\begin{table}[h!]
\centering
$\begin{array}{c|ccc}
\aut & & \t\text{'s} \\ 
\hline
\multirow{2}{*}{$D_2$} & 
\makebox(0,25){} 
\bpm z_1 & z_2\\ z_2 & z_1 \epm &
\bpm 2z_1 & z_1 \\ z_1 & z_2 \epm &
\dfrac12 \bpm z_1{-}z_2^{-1} & z_1{+}z_2^{-1}\\ z_1{+}z_2^{-1} & z_1{-}z_2^{-1}\epm 
\\[4mm]
& 
\dfrac12 \bpm z_1 & 1\\ 1 & z_2 \epm &
\bpm z_1 & \sqrt{1{+}z_1 z_2}\\ \sqrt{1{+}z_1 z_2} & z_2 \epm
\\[4mm] \hline
D_4 &
\makebox(0,25){} 
\dfrac12 \bpm z & 1\\ 1 & z \epm &
\dfrac14 \bpm 4z & 2z\\ 2z & z{-}4z^{-1} \epm &
\dfrac12 \bpm z{-}z^{-1} & z{+}z^{-1}\\ z{+}z^{-1} & z{-}z^{-1}\epm 
\\[4mm] \hline
D_6 & 
\makebox(0,25){} 
\dfrac12 \bpm 2z & z\\ z & 2z \epm &
\ \dfrac14 \bpm 4z & 2z{+}2\\ 2z{+}2 & \ z{-}3z^{-1}{+}2\epm &
\ \dfrac14 \bpm z{-}3z^{-1}{+}2 & z{+}3z^{-1} \\ z{+}3z^{-1} & z{-}3z^{-1}{+}2\epm 
\\[4mm] \hline
\Z_3{\rtimes} D_4 & &
\makebox(0,25){} 
\dfrac{i}{\sqrt3} \bpm 2 & 1\\ 1 & 2 \epm \\[4mm] \hline
\GL_2(3) & &
\makebox(0,25){} 
\dfrac13 \bpm 2i\sqrt2 \pm 1 & i\sqrt2 \mp 1\\ i\sqrt2 \mp 1 & 2i\sqrt2 \pm 1 \epm \\[4mm] \hline
\Z_{10} & &
\makebox(0,25){} 
\bpm \z_5 & \z_5+\z_5^3\\ \z_5+\z_5^3 & -\z_5^4 \epm &
\z_5 := e^{2\pi i/5}
\end{array}$
\caption{\label{tabGNU} 
$\Sp(4,\Z)$-equivalent forms of non-split moduli of rank-2 RSs with automorphisms from \cite{gottschling1961fixpunkte, gottschling1961fixpunktuntergruppen, Ueno1973}.}
\end{table}
The following $Sp(4,\Z)$ transformations demonstrate the equivalences of each row.
\begin{align}
G_T \circ G_R \circ 
\bpm 2z_1 & z_1\\ z_1 & z_2 \epm 
&=
\bpm z_2 & z_2{-}z_1\\ 
z_2{-}z_1 & z_2\epm, 
\label{spt1}\\
G_T \circ G_S \circ(1{\odot}S)\circ 
\dfrac12 \bpm z_1 & 1\\ 1 & z_2 \epm 
&=
\dfrac12 \bpm z_1{-}\frac1{z_2} & z_1{+}\frac1{z_2}\\ 
z_1{+}\frac1{z_2} & z_1{-}\frac1{z_2}\epm, 
\label{spt2}\\
(1{\odot}S) \circ 
\bpm z_1 & \sqrt{1{+}z_1 z_2}\\ \sqrt{1{+}z_1 z_2} & z_2 \epm
&=
\frac1{z_2} \, \bpm -1 & \sqrt{1{+}z_1 z_2}\\ \sqrt{1{+}z_1 z_2} & -1\epm, 
\label{spt3}\\
\nn\\
G_R \circ G_T \circ G_R \circ 
\frac14 \bpm 4z & 2z\\ 2z & z-\frac4z \epm 
&=
\frac12 \bpm 2z{-}\frac1{2z} & 2z{+}\frac1{2z}\\ 
2z{+}\frac1{2z} & 2z{-}\frac1{2z}\epm, 
\label{spt4}\\
(1{\odot}S) \circ G_S \circ 
\frac14 \bpm 4z & 2z\\ 2z & z-\frac4z \epm 
&=
\frac12 \bpm -\frac2z & 1\\ 
1 & -\frac2z\epm, 
\label{spt5}\\
\nn\\
G_R \circ G_T \circ G_R \circ T_{\bspm1&0\\0&0\espm} \circ
\frac14 \!\bpm 4z & 2z{+}2\\ 2z{+}2 & \ z{-}\frac3z{+}2\epm
&=
\frac14 \!\bpm z{-}\frac3z{+}2 & z{+}\frac3z \\ z{+}\frac3z & z{-}\frac3z{+}2\epm ,
\label{spt6}\\
G_R \circ T_{\bspm-1&0\\0&-1\espm} \circ (1{\odot}S) \circ G_S \circ
\frac14 \!\bpm z{-}\frac3z{+}2 & z{+}\frac3z \\ z{+}\frac3z & z{-}\frac3z{+}2\epm
&=
-\frac12 \left(1+\tfrac1z\right) \!\bpm 2 &1\\ 1 & 2\epm,
\label{spt7}\\
\nn\\
G_R \circ T_{\bspm1&0\\0&1\espm} \circ S \circ T_{\bspm1&0\\0&1\espm} \circ
\frac13 \!\bpm 2i\sqrt2{-}1 & i\sqrt2{+}1\\ i\sqrt2{+}1 & 2i\sqrt2{-}1 \epm
&=
\dfrac13 \!\bpm 2i\sqrt2{+}1 & i\sqrt2{-}1\\ i\sqrt2{-}1 & 2i\sqrt2{+}1 \epm .
\label{spt8}
\end{align}
In particular, the first and third entries in the $D_2$ row are the same up to a redefinition of $z_1$ and $z_2$, while \eqref{spt1}--\eqref{spt3} demonstrate the equivalence of the other three entries with the first, again up to a redefinition of $z_1$ and $z_2$.
\eqref{spt4} shows the equivalence of the second and third entries of the $D_4$ row, while \eqref{spt5} shows the equivalence of the second and first entries.
Note also that the specialization of \eqref{spt2} to $z_1=z_2$ directly shows the equivalence of the first and third entries of the $D_4$ row.
For the $D_6$ row, \eqref{spt6} shows the equivalence of the second and third entries, while \eqref{spt7} shows that of the third and first, up to a redefinition of $z$.
Finally, \eqref{spt8} demonstrates the equivalence of the two choices of sign in the $\GL_2(3)$ row.

These equivalencies can also be used to check the inclusions of modular subspaces shown in figure \ref{figA}.
In particular, $\cM_{D_2} \supset \cM_{D_4}, \cM_{D_6}$ since the first entries in the $D_4$ and $D_6$ rows of table \ref{tabGNU} are specializations of the first entry in the $D_2$ row.
$\cM_{D_4} \supset \cM_{\Z_3\rtimes D_4}$ since the $\Z_3\rtimes D_4$ modulus is the second entry in the $D_4$ row at $z=2i/\sqrt3$, and $\cM_{D_6} \supset \cM_{\Z_3\rtimes D_4}$ since the $\Z_3\rtimes D_4$ modulus is the first entry of the $D_6$ row also at $z=2i/\sqrt3$.
Finally, $\cM_{D_6} \supset \cM_{GL_2(3)}$ because $T_{\binom{1\ 0}{0\ 1}}$ acting on the $GL_2(3)$ entry with the bottom sign is the first entry of the $D_6$ row at $z=(2i\sqrt2-1)/3$.

\section{Automorphisms at rank 1}\label{appg1}

Any holomorphic family of genus 1 RSs together with a choice of its holomorphic 1-form can be presented as a binary-quartic plane curve%
\footnote{These are usually presented as families of degree-3 curves in $\P^2$, but we choose this presentation instead to ease the comparison to genus 2 RSs.}
\begin{align}\label{E1}
    y^2 &= c_4 x^4 + c_3 x^3 + c_2 x^2 + c_1 x + c_0 ,&
    \Om &= a \frac{dx}y ,
\end{align}
where the $c_n$ and $a$ coefficients vary holomorphically over some base space.
\eqref{E1} is projectivized as a degree 4 curve in the weighted projective space with homogeneous coordinates $[w:x:y] \in \P^2_{[1,1,2]}$.
The automorphisms of $\P^2_{[1,1,2]}$ are the group generated by the linear and quadratic coordinate transformations,
\begin{align}\label{E2}
    \bspm x \\ w\espm &\mapsto G \bspm x \\ w\espm, &
    & \bspm G^1_1 & G^1_2 \\ G^2_1 & G^2_2 \espm \doteq G 
    \in \GL(2,\C),
    \nn\\
    y &\mapsto Q_0 y + Q_{11} x^2 + Q_{12} xw + Q_{22} w^2, &
    & \, Q_0 \in \C^*, \quad Q_{ij} \in \C .
\end{align}
But for a general (non-singular) holomorphic family of degree 4 curves, the $Q$ coordinate transformations can be used to put the curve into the binary-quartic form \eqref{E1} with $Q_{ij}$ which are holomorphic (in fact, linear) functions of the coefficients of the degree 4 curve.  Furthermore, the $Q$ coordinate transformation freedom is completely fixed in this way.
Thus, without loss of generality, we can present any holomorphic family of genus 1 RSs as in \eqref{E1}.

The remaining $\GL(2,\C)$ reparameterizations act in the $w=1$ patch as
\begin{align}\label{E3}
x &\mapsto \frac{G^1_1 x+ G^1_2}{G^2_1 x+G^2_2} , &
y &\mapsto \frac {y}{(G^2_1 x+ G^2_2)^2} ,&
\Om &\mapsto (\det G) \, \Om,
\end{align}
and preserve the binary-quadratic form of the curve.
When the transformation is in the $\Z_2$ subgroup generated by $-I_2$ in the center of $\GL(2,\C)$, it acts trivially on the points of the curve and the 1-form.
So the nontrivial coordinate transformations are only those in $\GL(2,\C)/\Z_2$.

Automorphisms of a genus-1 RS are then $\GL(2,\C)/\Z_2$ coordinate transformations which preserve the values of the $c_n$ coefficients.
Genus-1 RSs have automorphism groups 
\begin{align}\label{g1aut}
    \aut = (\U(1)\times\U(1))\rtimes \bar\aut,
\end{align}
where the discrete subgroups, $\bar\aut$, and their $\GL(2,\C)/\Z_2$ actions on representative curves are given in table \ref{tabB}.  Here $\t$ is the complex modulus of the genus 1 RS chosen in the standard fundamental domain in the complex upper half plane $\cH_1 \doteq \{z \, |\, \Im z >0\}$ of the moduli space of genus 1 RSs, $\cM_1 = \cH_1/\Sp(2,\Z)$.
In particular, all genus 1 RSs have the ``hyperelliptic" $\Z_2$ automorphism factor, which is enhanced for two special genus 1 RSs as shown.

\begin{table}[h!]
\centering
$\begin{array}{c|c|c|c}
\ \bar\aut\ \ &c^\aut(x) &\ \bar\aut\subset B(2,\C)/\Z_2 \ \ 
&\t \\ [1mm]
\hline \makebox(0,15){}
\Z_2  & \ x^3 {+} c_1 x {+} c_0\ \ 
& \big\langle e^{i\pi/2}\bspm 1 &\\&1\espm \big\rangle 
& \cM_1 \\[2mm]
\Z_4 & x(x^2{-}1)
& \big\langle e^{i\pi/4}\bspm 1 &\\&-1\espm \big\rangle 
& e^{i\pi/2} \\[3mm]
\Z_6 & x^3{-}1
& \big\langle e^{i\pi/6}\bspm 1&\\& e^{i\pi 4/3}\espm \big\rangle 
& \ e^{i\pi/3} \ \ 
\end{array}$
\caption{\label{tabB} 
Some data describing the automorphism groups of genus 1 RSs.}
\end{table}

The continuous normal subgroup of $\U(1)\times\U(1)$ automorphisms are the doubly-periodic translations of the torus.
Their corresponding $\GL(2,\C)$ transformations have a famously complicated algebraic expression (closely related to the group law of an elliptic curve).
Since translations do not change a choice of 1-cycle basis on the torus, a twist of a genus 1 family by these automorphisms does not change the CB effective action.
Thus we can safely ignore the $\U(1)\times\U(1)$ automorphisms.

But, because $\bar\aut$ is not normal in $\aut$, in order to pick an embedding of $\bar\aut$ in $\GL(2,\C)/\Z_2$, we must further restrict the form of the curve so as to fix the translation symmetry.
This is done by fixing the coordinates of one of the points of the curve for the whole family, i.e., by making it a family of elliptic curves. This is traditionally done by fixing one of the branch points of \eqref{E1} at $x=\infty$, thereby making it a binary-cubic curve.
The subgroup of $\GL(2,\C)/\Z_2$ which fixes $x=\infty$ is the Borel subgroup of upper triangular matrices, $B(2,\C)/\Z_2 \subset \GL(2,\C)/\Z_2$.
We can further fix the Borel subgroup to a diagonal subgroup by putting the curve into Weierstrass form
\begin{align}\label{E6}
    y^2 &= x^3 + c_1 x + c_0 ,
\end{align}
which leaves unfixed only reparameterizations of the form
\begin{align}\label{BWeier}
    B = \bpm \b^{-1} & \\ & \b^3 \epm, \qquad \b\in\C^*.
\end{align}
They act on the coefficients and 1-form as
\begin{align}\label{BWeier2}
    B: c_0 &\mapsto \b^{12} c_0, & 
    c_1 &\mapsto \b^{8} c_1, & 
    \Om &\mapsto \b^2 \Om .
\end{align}
The ones which leave the coefficients of the curve invariant are those
with $\b^4=1$ if $c_0c_1\neq0$, or $\b^8=1$ if $c_0=0$, or $\b^{12}=1$ if $c_1=0$, giving the classification in table \ref{tabB}.

Finally, the remaining \eqref{BWeier} reparameterization freedom can be fixed by putting the curve and 1-form into ``Weierstrass-canonical" frame
\begin{align}\label{E7}
    y^2 &= x^3 + c_1 x + c_0, & \Om &= \frac{dx}y .
\end{align}
It is straight forward to check that a $B(2,\C)/\Z_2$ transformation exists that can put a curve \eqref{E6} and general 1-form into this form everywhere on a cut CB, and to check that the only $B(2,\C)/\Z_2$ transformation which both preserves the $c_n$ coefficients of the Weierstrass-canonical curve form as well as fixes $\Om=dx/y$ is the identity.
Thus, the Weierstrass-canonical frame \eqref{E6} completely fixes the remaining reparametrization freedom.
Furthermore, since the family of RSs is holomorphic on CB$^*$, in this frame the coefficient functions will be meromorphic (i.e., single-valued) on the CB.
This Weierstrass-canonical frame is the usual starting point for deriving the familiar Kodaira classification of possible rank 1 scale-invariant CB geometries, see, e.g., \cite{Minahan:1996fg, Minahan:1996cj}.

Suppose that we have a genus 1 SW curve in Weierstrass form varying holomorphically over a scale-invariant rank 1 CB$^*$ but it is not in Weierstrass-canonical frame, i.e., its 1-form $\Om = a(u) dx/y$ is not necessarily of canonical form.
Then, as we have seen, this data must be supplemented by an automorphism twist,
\begin{align}
    \a : \pi_1(\CB^*) \to \bar\aut .
\end{align}
associated to the possible non-single-valuedness of the 1-form coefficient function $a(u)$. 
We will now describe this twist explicitly.

Assuming a freely generated CB chiral ring, the CB is $\C$ as a complex manifold; let $u$ be a complex coordinate on the CB.
Superconformal invariance implies a holomorphic $\C^*$  action on the CB, and let $u$ be a complex coordinate linearizing this action.   
Then $u$ has scaling dimension $\D_u>0$, and the geometry of the CB is thus that of a flat cone with tip at $u=0$.
Since the only singularity as at the tip, $\CB^*\simeq \C^*$, and $\pi_1(\CB^*) \simeq \Z$, generated by a simple closed path $\g$ encircling the tip once counterclockwise.
Take the cut CB to be CB$^*$ minus the positive real axis.
The automorphism twist is then specified by the value of $\a(\g)$ in $\bar\aut$.  

The representation $N: \bar\aut \to \Sp(2,\Z)$ depends on the choice of 1-cycle basis through the value of the complex modulus of the curve, $\t\in\cH_2$.
Choosing, as in table \ref{tabB} $\t=i$ and $\t=e^{i\pi/3}$ for the curves with $\bar\aut=\Z_4$ and $\z_6$, respectively, it is then easy to see that the image of $N$ in $\Sp(2,\Z)$ for the possible automorphism groups are
\begin{align}\label{Nbaraut}
     \im N(\Z_2) &= \big\langle -I \big\rangle,&
     \im N(\Z_4) &= \big\langle S \big\rangle,&
     \im N(\Z_6) &= \big\langle TS \big\rangle, 
\end{align}
where $S \doteq \bspm 0 & -1\\1&0\espm$ and $T \doteq \bspm 1 & 1\\ 0 & 1\espm$ generate $\Sp(2,\Z)$ and satisfy $S^2=(TS)^3=-I$.

A SW curve with automorphism twist $\a(\g)$ and associated $\Sp(2,\C)$ representative $N(\a(\g)) \doteq \bspm A&B\\C&D\espm$, will also have an associated $B_{\bar\aut} \in B(2,\C)/\Z_2$ coordinate transformation which preserves the Weierstrass form of the curve, so is of the form \eqref{BWeier}.
$B$ is determined in terms of $N$ by the same argument using the invariance of the period matrix that was used to deduce \eqref{ad3} in the rank 2 case.
In the rank 1 case this implies that 
\begin{align}\label{Bbaraut}
   ( \det B_{\bar\aut} )^{-1} = C\t+D , 
\end{align}
giving $B_{\bar\aut}$ as in \eqref{BWeier} with
\begin{align}\label{Bbaraut2}
    \b=(C\t+D)^{-1/2}.
\end{align}
Note that the sign of the square root is immaterial since it is in the $\Z_2$ center factor.
Note also that, in contrast to the rank 2 case, the dependence of $B_{\bar\aut}$ on the electric period $\cA_\w$ drops out in the rank 1 case.
We evaluate $\b$ in \eqref{Bbaraut2} for each $N$ in $\bar\aut$ using the corresponding value of $\t$ to find 
\begin{align}\label{Bbaraut3}
    \b &= e^{i\pi m/n} &
    &\text{if}& \a(\g) &= m \in \Z_n \doteq \bar\aut,
\end{align}
which simply reflects the phase monodromy associated with the flat conical CB geometries occurring in the Kodaira classification.

\bibliographystyle{Auxiliary/JHEP}

\end{document}